\documentclass[screen,nonacm]{acmart}
\usepackage{amsmath,amsfonts}
\usepackage{algorithmic}
\usepackage{algorithm}
\usepackage{array}
\usepackage[caption=false,font=normalsize,labelfont=sf,textfont=sf]{subfig}
\usepackage{textcomp}
\usepackage{stfloats}
\usepackage{url}
\usepackage{verbatim}
\usepackage{graphicx}
\usepackage{xcolor}
\usepackage{tabularray}
\usepackage[utf8]{inputenc}
\usepackage{tcolorbox}
\usepackage{hyperref}
\usepackage{pgfplots}
\usepackage{standalone}
\usepackage{pgf-pie}
\usepackage[T1]{fontenc}
\usepackage{multicol}
\UseTblrLibrary{diagbox}
\usepackage{makecell}
\usepackage{enumitem}

\pgfplotsset{compat=1.18}
\usepgfplotslibrary{colorbrewer}
\hypersetup{
    pdfborder={0 0 0}
}

\SetTblrStyle{caption}{font=\footnotesize}
\SetTblrTemplate{lasthead}{empty}

\AtBeginDocument{%
  }

\begin{document}

\title{LLM-Assisted Empirical Software Engineering: Systematic Literature Review and Research Agenda}

\author{Victória Gomes}
\email{oliveiragov@vcu.edu}
\affiliation{%
  \institution{Virginia Commonwealth University}
  \country{USA}
}

\author{Delaney Selb}
\email{selbdm@vcu.edu}
\affiliation{%
  \institution{Virginia Commonwealth University}
  \country{USA}
}

\author{Fabio Palomba}
\email{fpalomba@unisa.it}
\affiliation{%
  \institution{University of Salerno}
  \country{Italy}
}

\author{Rodrigo Spinola}
\email{spinolaro@vcu.edu}
\affiliation{%
  \institution{Virginia Commonwealth University}
  \country{USA}
}

\author{David Lo}
\email{davidlo@smu.edu.sg}
\affiliation{%
  \institution{Singapore Management University}
  \country{Singapore}
}

\renewcommand{\shortauthors}{Gomes et al.}

\begin{abstract}
\textbf{\textit{Context:} }
Empirical Software Engineering (ESE) faces increasing challenges due to data scale, methodological complexity, and reproducibility concerns. Large Language Models (LLMs) have emerged as promising tools to support empirical workflows, yet their use remains fragmented, with no comprehensive synthesis to guide responsible adoption.

\textbf{\textit{Aims:} }
This study analyzes how LLMs are used in ESE, examining supported tasks, phases of the empirical lifecycle, integration into workflows, reported benefits and limitations, and the extent of reproducibility-related reporting. It also identifies gaps and future research directions.

\textbf{\textit{Method:}} 
We conducted a systematic literature review of peer-reviewed papers (2020-2025) across 12 leading software engineering venues, resulting in 50 primary studies analyzed through qualitative and quantitative synthesis.

\textbf{\textit{Results:} }
We identified 69 LLM-assisted tasks, mainly in mining software repositories and controlled experiments, focusing on classification, filtering, and evaluation. LLMs are used across multiple phases but are concentrated in data processing and analysis. Their integration is largely automation-oriented, with limited decision-support use. Benefits emphasize efficiency and scalability, while limitations include hallucinations, inconsistency, prompt sensitivity, and reproducibility issues. Reporting practices are often incomplete.

\textbf{\textit{Conclusion:} }
LLM use in ESE is growing but remains automation-driven, with gaps in human-centered integration and transparency. We outline implications and research agenda for responsible use.
\end{abstract}







\maketitle

\section{Introduction}
\label{sec:intro}

Empirical Software Engineering (ESE) provides the methodological foundation for producing rigorous, evidence-based knowledge about software systems, development practices, tools, and human aspects of software engineering. It relies on a range of empirical methods, including controlled experiments, case studies, surveys, mining studies, and secondary studies, to investigate how software is developed and maintained in real-world settings, as well as how methods and technologies affect quality, productivity, and sustainability. Over the past two decades, the field has established widely accepted methodological guidelines for conducting and reporting empirical research, particularly in experimentation, case study research, and systematic literature reviews (SLRs) \cite{kitchenham2002preliminary,ExperimentationSEWohlin,runeson2009guidelines,petersen2008systematic}.

At the same time, ESE faces increasing challenges related to scale, complexity, and methodological rigor. Contemporary software systems generate massive and heterogeneous data sources, including source code repositories, issue trackers, pull requests, developer communications, and gray literature, that demand large-scale analysis and often exceed the practical capacity of purely manual or traditional semi-automated techniques \cite{Hassan2008,Shull2008}. In parallel, qualitative components of empirical research, such as coding interviews, analyzing open-ended survey responses, and conducting evidence synthesis, remain particularly labor-intensive and dependent on interpretive judgment, making them susceptible to researcher bias and inconsistency \cite{Shull2008}. Recent discussions in ESE have further highlighted reproducibility challenges stemming from complex datasets, evolving tools, and insufficient reporting of analytic procedures \cite{Collberg2016}. Together, these pressures raise broader concerns regarding efficiency, reproducibility, and the long-term sustainability of empirical research workflows \cite{ExperimentationSEWohlin}.

\subsection{Background and Motivation}

As empirical datasets continue to grow and methodological expectations for transparency and rigor increase, there is a pressing need for scalable and reliable forms of computational assistance that can support, rather than replace, human researchers in conducting empirical studies. Recent advances in Large Language Models (LLMs) offer new opportunities to address these challenges. Modern LLMs demonstrate strong capabilities in natural language understanding, summarization, classification, reasoning, and code-related tasks, as evidenced by their performance across a wide range of natural language processing benchmarks and software engineering applications \cite{brown2020language,wei2022chain,Chen2021,LLM4SE_LITREVIEW}. Prior secondary studies have documented the rapid adoption of LLMs across software engineering activities, including code generation, testing, program repair, and documentation \cite{LLM4SE_LITREVIEW,zhou2022survey}. 

Beyond serving as artifacts to be evaluated, LLMs are increasingly being used as \emph{research instruments} to support ESE activities. Emerging work shows that LLMs can assist with information extraction from software repositories, qualitative coding of interviews and surveys, large-scale evidence synthesis, and study screening and classification in systematic reviews \cite{LLMs4SelectionPhaseInSLR,LLMs4QualAnalysisInSE,LLMs4MSR}. Preliminary evidence suggests that such uses can improve scalability and reduce researcher effort, while also introducing new risks related to hallucinations, bias, lack of transparency, and threats to validity.

Despite growing interest, several considerations still limit a clear understanding of how LLMs are used in ESE. First, research on LLM-assisted ESE is scattered across different venues, due to the emerging and widespread adoption of LLMs in studies targeting diverse tasks and methodological contexts, as well as the lack of consolidated evidence describing how LLM-assisted ESE has been performed. This makes the evidence base harder to identify, compare, and consolidate. In addition, many studies focus on isolated applications, i.e., they investigate the use of LLMs within a single task, dataset, or methodological setting, making it difficult to determine whether their findings generalize beyond specific contexts. Validation strategies also vary substantially across studies, which complicates cross-study comparison and limits confidence in the robustness of reported results. Moreover, benefits and limitations are often reported unevenly, making it harder to build a cumulative understanding of when and how LLMs can be used effectively. As a result, it remains difficult to assess where LLMs can be reliably integrated into empirical workflows, what methodological safeguards are needed, and which research gaps remain open.

\subsection{Problem Statement}

Although LLMs have been widely studied in software engineering and are increasingly adopted in empirical research workflows, there is currently no systematic synthesis of how these models are used to support empirical software engineering research activities. In particular, it remains unclear which research tasks and phases are supported, how LLMs are integrated into empirical processes, and what benefits, limitations, and reporting practices characterize their use.

Consequently, fundamental questions about the current state of practice remain unanswered: Which ESE research tasks and phases are supported by LLMs? How are LLMs integrated into empirical research workflows? What benefits and limitations are reported when LLMs support empirical activities?  To what extent do studies report the information required to enable the reproducible use of LLMs? Without a structured synthesis of existing evidence, researchers and practitioners lack a consolidated understanding of how LLMs are being used and how their use should be critically assessed in empirical software engineering.

Recent survey papers in the broader LLM4SE literature have provided important foundations, but none addresses this gap directly. Hou et al. \cite{LLM4SE_LITREVIEW}, for example, present a broad systematic literature review of LLMs for software engineering, covering 395 studies and synthesizing models, datasets, optimization strategies, evaluation approaches, and SE tasks across the software lifecycle. Their review offers an SE-wide perspective, but it does not specifically examine the use of LLMs in empirical software engineering research workflows. Zadenoori et al. \cite{zadenoori2025largelanguagemodelsllms} narrow the focus to requirements engineering (RE), reviewing 74 primary studies and examining supported RE activities, prompting strategies, evaluation methods, and resources, but their analysis remains restricted to a single SE subdomain. He et al. \cite{LLMAsAJudge4SE}, in turn, focus on a specific emerging paradigm by reviewing 42 studies on LLM-as-a-Judge in software engineering, examining the use of LLMs as evaluators of software artifacts and outlining the limitations and future directions of that paradigm.

In parallel, recent work has started to propose empirical guidelines for the use of LLMs in software engineering research (e.g., \cite{baltesllm_guidelines}), highlighting the need for transparent, rigorous, and reproducible practices. This effort is essential to standardizing the design, execution, and reporting of LLM-assisted empirical studies; however, it does not provide a consolidated view of how such practices are currently adopted in empirical workflows. As a consequence, it serves a complementary role to our work, which aims to synthesize and characterize their actual use in the literature.

\subsection{Objectives and Research Questions}

The primary goal of this study is to understand how LLMs are used to support ESE research and what implications this has for research rigor and practice. To frame this goal, we define the following overarching research question:

\begin{tcolorbox}[title=Overarching Research Question, colframe=black, fontupper=\small, fonttitle=\small\bfseries]
How are large language models used to support empirical software engineering research, and what benefits and limitations arise from their use?
\end{tcolorbox}

This overarching research question is decomposed into five more fine-grained research questions that examine complementary dimensions of LLM-assisted ESE. Specifically, this study investigates: (i) which ESE research tasks are supported by LLMs, (ii) which ESE research phases are supported by LLMs, (iii) how LLMs are operationalized within empirical research workflows, (iv) the benefits and limitations reported when using LLMs in empirical research activities, and (v) the extent to which these studies report the information required to enable reproducible LLM usage. The detailed formulation and justification of these research questions are presented in Section~\ref{sec:methodology}.

\subsection{Study Overview}

To systematically investigate the role of LLMs in ESE research, we conducted a systematic literature review (SLR) following established methodological guidelines for secondary studies in software engineering \cite{Kitchenham2007,ExperimentationSEWohlin}. The review examined peer-reviewed publications published between 2020 and 2025 across 12 leading software engineering conferences and journals. Through a multi-stage screening and selection process, 50 primary studies were ultimately identified for in-depth data extraction and synthesis. These studies constitute the empirical foundation of this review.

\subsection{Contributions}

This paper makes the following contributions:

\begin{itemize}
    \item A taxonomy of ESE research tasks supported by LLMs, identifying how LLMs are applied across different phases of empirical research;
    \item An analysis of how LLMs are operationalized within empirical research workflows, describing integration patterns and usage strategies;
    \item A synthesis of reported benefits and limitations of using LLMs to support ESE research;
    \item An assessment of reporting practices for reproducible LLM usage, examining whether studies disclose essential information such as model versions, prompts, and configuration details;
    \item A discussion of research gaps and methodological recommendations to guide future work on the responsible integration of LLMs into ESE research;
    \item A discussion of new frontiers for further scaling up the use of LLMs in ESE research to facilitate greater automation and discovery.
\end{itemize}

\subsection{Paper Organization}

The remainder of this paper is organized as follows. 
Section~\ref{sec:background} introduces the background concepts. Section~\ref{sec:methodology} describes the research methodology, including the study design and data collection procedure. 
Section~\ref{sec:overview} provides an overview of the primary studies included in the review. 
Section~\ref{sec:dataAnalysis} details the data analysis approach adopted to address the research questions. 
Section~\ref{sec:results} presents the results for each research question. 
Section~\ref{sec:discussion} discusses and synthesizes these findings. 
Section~\ref{sec:researchAgenda} outlines a research agenda by identifying key gaps in the literature and proposing directions for future work. 
Section~\ref{sec:threats} addresses threats to validity, Section~\ref{sec:relatedwork} reviews related work, and Section~\ref{sec:conclusion} concludes the paper.

\section{Background}
\label{sec:background}

ESE encompasses a wide range of study designs, including controlled experiments, case studies, surveys, ethnographic studies, and secondary studies. These methods are governed by established guidelines that stress validity assessment, replicability, and explicit reporting of assumptions and limitations \cite{kitchenham2002preliminary,Kitchenham2007}. Despite methodological advances, many empirical studies remain labor-intensive, particularly when handling large datasets, unstructured textual artifacts, or qualitative evidence, motivating continued interest in computational support for empirical research.

Automation has long been used to support empirical studies in software engineering. Techniques such as mining software repositories (MSR), static and dynamic analysis, and natural language processing (NLP) have enabled large-scale analyses of source code, issue trackers, version control systems, and developer communications \cite{Hassan2008,Zimmermann2005,codabux2024teaching}. Machine-learning approaches have further expanded this capability, supporting tasks such as defect prediction, effort estimation, and topic modeling \cite{Hindle2012}.

While these approaches have significantly improved scalability, they typically rely on task-specific models, handcrafted features, and domain expertise. Moreover, traditional automation often struggles with activities requiring semantic interpretation, abstraction, or contextual reasoning, capabilities central to qualitative analysis, theory building, and evidence synthesis in ESE. These limitations have prompted interest in more general and flexible computational models that can assist researchers across diverse empirical tasks.

\subsection{Large Language Models}
LLMs are general-purpose language processing systems trained on extensive textual corpora to learn broad statistical regularities in language \cite{brown2020language}. Rather than being engineered for a single predefined task, LLMs acquire flexible capabilities that enable them to generate, interpret, summarize, classify, and transform textual and code-related artifacts across domains \cite{brown2020language,Bommasani2021}. This task-agnostic nature distinguishes LLMs from earlier task-specific machine learning approaches and positions them as adaptable computational assistants capable of supporting diverse knowledge-intensive activities \cite{Bommasani2021}.

Current LLMs have demonstrated strong performance across a wide range of tasks, including summarization, classification, question answering, reasoning, and code-related activities \cite{brown2020language,Chen2021}. In software engineering, they have been shown to support activities such as code and test generation, documentation, program repair, and requirements analysis \cite{Chen2021,Pearce2022}. Unlike earlier NLP or machine-learning approaches that typically required explicit feature engineering and narrowly scoped models, LLMs can operate directly on raw artifacts, making them particularly attractive for tasks involving heterogeneous or unstructured data, conditions commonly encountered in ESE studies.

\subsection{LLM Assistance Across the Empirical Research Lifecycle}
The use of LLMs in empirical research represents a shift from task-specific automation toward assistive intelligence, in which models collaborate with human researchers rather than replace them. This perspective aligns with broader research on human-AI collaboration, which emphasizes human oversight, interpretability, and shared control over decision-making \cite{Amershi2019}. 

Within the context of ESE, LLMs have been explored as aids for activities such as qualitative coding, study classification, data labeling, thematic synthesis, and narrative summarization \cite{Gilardi2023}. These applications suggest that LLMs can reduce cognitive load and accelerate empirical workflows, particularly during exploratory phases or when handling large volumes of textual evidence. However, their probabilistic and opaque nature also introduces risks related to hallucinations, bias, reproducibility, and the influence of model outputs on human judgment, concerns that are especially critical in empirical research settings.

To systematically reason about the role of LLMs in ESE, it is useful to situate their use within the empirical research lifecycle, which typically includes problem formulation, study design, data collection, analysis, synthesis and interpretation, and reporting. Prior work has shown that computational tools can affect different stages of this lifecycle in distinct ways \cite{Shull2008,ExperimentationSEWohlin}. For instance, LLMs may contribute to early-stage activities, such as refining research questions or identifying constructs, as well as later-stage tasks including qualitative coding, evidence aggregation, and reporting \cite{Gilardi2023}. Importantly, the degree of autonomy assigned to LLMs varies across studies. In some cases, model outputs are treated as preliminary suggestions subject to careful human validation, whereas in others they assume a more central analytic role \cite{Weidinger2022,Bender2021}. These variations have direct implications for internal validity, construct validity, and transparency, underscoring the need for a structured synthesis of how LLMs are integrated into empirical research.

\section{Research Methodology}
\label{sec:methodology}

In this work, we conducted an SLR to examine how LLMs have been used to support empirical research. To ensure comprehensive and transparent reporting of the study’s objectives, procedures, and findings, we followed a well-defined review protocol grounded in established guidelines for empirical research~\cite{EBSEKitchenham, ExperimentationSEWohlin}. These guidelines promote methodological rigor and transparency, systematically guiding the planning, execution, and reporting phases of the review.

This section first introduces the terminology and definitions adopted in this study (\ref{subsec:terminologyDefinitions}). It then describes the research questions and their rationale (\ref{subsec:researchQuestions}), the inclusion and exclusion criteria for study selection (\ref{subsec:InclusionExclusionCriteria}), the search strategy (\ref{subsec:searchStrategy}) and study selection processes (\ref{subsec:studySelection}), as well as the data extraction (\ref{subsec:dataExtraction}).

\subsection{Terminology Clarification and Scope of LLM Usage}
\label{subsec:terminologyDefinitions}

\textbf{Terminology Clarification.} To clarify the terminology used in the research design of this study, we adopt a hierarchical structure that distinguishes different levels of activity, inspired by how these concepts are used in Wohlin et al.~\cite{ExperimentationSEWohlin}, although the authors do not provide explicit definitions. Within this structure, a \textbf{phase} represents a macro-level grouping in the research lifecycle (e.g., planning \& design, data collection, data processing, analysis \& synthesis, and reporting). Each phase can be further decomposed into \textbf{steps}, corresponding to discrete, method-specific procedures. In turn, each step comprises \textbf{tasks}, which we interpret as fine-grained functional activities performed during that step. 

To illustrate this hierarchy, consider an SLR workflow. At the highest level, the \textit{data processing} phase includes the \textit{screening} step. This step is operationalized through multiple tasks, such as (i) title-abstract screening, where irrelevant studies are excluded based on predefined criteria, and (ii) full-text screening, where eligibility is assessed in detail. If LLMs are used in this context, they may support specific tasks (e.g., automatically classifying papers as relevant or not), while the step (screening) and phase (data processing) provide the broader methodological context in which this support occurs.

\textbf{Scope of LLM Usage.} In this study, we include papers in which LLMs are used either to support empirical software engineering research or to be empirically studied within that context. In the first case, LLMs directly contribute to research activities such as data collection, data processing, analysis, synthesis, or reporting. In the second case, LLMs are themselves the focus of the paper, but the study is included only when it examines their role, behavior, or performance in relation to a software engineering research activity. Therefore, we exclude studies that treat LLMs merely as development artifacts or evaluate them only through standalone technical benchmarks, such as code generation or testing tasks, unless these evaluations are explicitly connected to an empirical software engineering research process.

\subsection{Research Questions}
\label{subsec:researchQuestions}

To guide our investigation, we defined five research questions (see Table \ref{tab:research-questions}). In \textbf{RQ1} we investigate which ESE research tasks are currently supported by LLMs. Identifying these tasks is essential for mapping the current landscape of LLM-assisted empirical work and for understanding how such models are operationalized in practice. By classifying the tasks supported, such as data extraction, qualitative coding, screening, or synthesis, we provide a structured view of how LLMs are applied across empirical research activities.

While RQ1 focuses on fine-grained tasks, \textbf{RQ2} shifts the analysis to a higher level of abstraction, enabling us to understand where in the research lifecycle LLMs are being integrated. This perspective is important to assess whether LLM adoption is concentrated in specific phases (e.g., data collection and analysis) or distributed across the full research process, from study design to reporting. By mapping LLM-supported tasks to established ESE research phases, we can identify patterns of use, gaps in support, and opportunities for broader integration.

\begin{table}[ht]
\caption{List of research questions addressed in the study}
\label{tab:research-questions}
\begin{tblr} {
    colspec = {Q[c, m, 0.5cm] X[j, m]},
    hlines,
    row{1} = {bg=black, fg=white, font=\bfseries},
    hline{7} = {1pt, solid},
    rows = {font=\footnotesize},
    rowsep = 1pt
}
    \textbf{Id} & \textbf{Research question} \\
    RQ1 & What empirical software engineering research tasks are supported by LLMs in LLM-assisted research studies? \\

    RQ2 & Which phases of the ESE research process are supported by LLMs in LLM-assisted studies? \\

    RQ3 & How are LLMs integrated into empirical software engineering research workflows? \\
    
    RQ4 & What benefits and limitations are reported when using LLMs to support empirical software engineering research tasks? \\
    
    RQ5 & To what extent do studies proposing LLM-assisted empirical software engineering research report the information required to enable reproducible use of LLMs? \\
\end{tblr}
\end{table}

Next, in \textbf{RQ3}, we examine how LLMs are integrated into ESE research workflows. Building on the tasks identified in RQ1 and their distribution across research phases in RQ2, this question shifts the focus to the nature of LLM involvement in these activities. Specifically, we characterize different modes of integration by distinguishing whether LLMs act as mechanisms for automation, augmentation, decision support, or evaluation. This analysis allows us to move beyond where LLMs are used to understand how they are used, revealing patterns of human-AI collaboration and the extent to which LLMs replace, support, or complement human effort. By doing so, we provide a comprehensive view of how LLMs are shaping empirical research workflows and identify opportunities for more balanced and collaborative integration strategies.

In \textbf{RQ4}, we analyze the benefits and limitations reported in the literature when LLMs are used to support ESE tasks. Synthesizing both advantages and drawbacks provides critical insight into the practical and methodological impact of LLM adoption. Understanding these trade-offs is necessary for assessing the reliability of LLM-assisted workflows and for informing the responsible and rigorous integration of such models into empirical research.

Finally, in \textbf{RQ5}, we examine how studies proposing LLM-assisted ESE research tasks report the experimental settings (e.g., prompts, model versions, parameters, and configurations) under which their experiments were conducted, since clear and comprehensive reporting of these settings is essential to ensure reproducibility when LLMs are involved.

\subsection{Inclusion and Exclusion Criteria}
\label{subsec:InclusionExclusionCriteria}
Before performing the screening step, a set of selection criteria was established to guide the systematic identification of relevant studies, as shown in Table~\ref{tab:inclusion-exclusion-criteria}. These criteria operationalize the scope of the review by distinguishing between studies that use LLMs as methodological support within ESE research workflows and those that apply LLMs to general software development tasks unrelated to the conduct of empirical research. Therefore, in particular, the review focuses on LLMs as research-support tools rather than as general-purpose development assistants.

\begin{table}[ht]
\caption{Inclusion and Exclusion criteria}
\label{tab:inclusion-exclusion-criteria}
\begin{tblr}{
    colspec = {Q[c, m, 1.2cm] X[j, m] Q[c, m, 1cm] Q[c, m, 1cm]},
    hlines,
    row{1,5} = {bg=black, fg=white, font=\bfseries},
    hline{9} = {1pt, solid},
    rows = {font=\footnotesize},
    rowsep = 1pt
}
    \SetCell[c=2]{l} \textbf{Inclusion criteria} &  \\
    IC1 & Studies in which LLMs are used to support, automate, or empirically examine activities within empirical software engineering research workflows. This includes studies where LLMs assist research tasks such as study screening, data extraction, qualitative coding, evidence synthesis, or research reporting, as well as studies that evaluate the role, behavior, or performance of LLMs in relation to an ESE research activity \\
    IC2 & Peer-reviewed papers published in journals or conference proceedings \\
    IC3 & Studies explicitly situated within the field of software engineering research \\
    \SetCell[c=2]{l} \textbf{Exclusion criteria} &  \\
    EC1 & Studies that do not use LLMs to support or automate ESE research tasks. This includes studies that apply LLMs exclusively to software development tasks unrelated to the conduct of empirical research. \\
    EC2 & Non-peer-reviewed publications (e.g., editorials, prefaces, posters, opinion pieces, blog posts, or technical reports without peer review) \\
    EC3 & Papers not written in English \\
\end{tblr}
\end{table}

\subsection{Search Strategy}

\label{subsec:searchStrategy}
We initially explored the possibility of defining a database search string that would precisely retrieve studies aligned with IC1. However, this approach proved impractical. Broad search terms related to LLMs (e.g., ``LLM'', ``Generative AI'', ``ChatGPT'') consistently returned a large volume of studies applying LLMs to general software development tasks (e.g., code generation, testing, or documentation), which fall outside the scope of this review. Conversely, more restrictive keyword combinations failed to reliably capture studies in which LLMs were used to support empirical software engineering (ESE) research tasks, since such usage is frequently described within methodological sections rather than emphasized in titles, abstracts, or author-provided keywords.

Given this trade-off between recall and precision, and the difficulty of operationalizing IC1 through automated keyword-based filtering alone, we adopted a curated corpus construction strategy based on a set of leading empirical software engineering venues (see Table~\ref{tab:venue-list}). To construct the dataset, we manually downloaded the full proceedings of these venues and examined them to identify studies focused on LLM-supported ESE research tasks.

This design choice is supported by prior research on systematic literature studies in software engineering, which highlights the limitations of relying exclusively on keyword-based database searches. In particular, previous work has shown that database searches may suffer from low precision and recall, especially in emerging areas where terminology is not yet stable or where relevant information is embedded in methodological sections rather than explicitly stated in titles, abstracts, or keywords~\cite{jalali2012systematic, mourao2020performance, wohlin2022successful}. In such contexts, alternative or complementary strategies, including snowballing and curated source selection based on high-quality venues, have been recommended to improve the coverage and relevance of identified studies~\cite{wohlin2014guidelines}. Following this rationale, we restricted our corpus to well-established venues that are known to publish rigorous empirical software engineering research. This restriction made the analysis feasible, while ensuring that the collected studies are both relevant and of high quality.

Venue selection was guided by objective quality and scope criteria: we included well-established, peer-reviewed venues that (i) are classified as CORE A*/A (or equivalent high-impact rankings), (ii) are widely recognized as primary publication outlets in software engineering, and (iii) explicitly publish empirical software engineering research. These criteria ensure coverage of the principal venues where rigorous methodological contributions in ESE are most likely to appear, while maintaining quality control and reproducibility of the search process. Both conference and journal venues were considered to ensure comprehensive coverage of empirical software engineering research across different publication types, capturing both mature journal contributions and emerging conference studies.

Finally, we restricted the analysis to publications between 2020 and 2025, reflecting the recent emergence of LLM applications in this domain.
\begin{table}[ht]
    \caption{List of venues used as a source for the SLR.}
    \label{tab:venue-list}
    \begin{tblr} {
        colspec = {Q[c, m, 1.5cm] X[j, m] Q[c, m, 1cm]},
        hlines,
        row{1} = {bg=black, fg=white, font=\bfseries},
        hline{14} = {1pt, solid},
        rows = {font=\footnotesize},
        rowsep = 1pt
    }
        \textbf{Venue type} & \textbf{Name} & \textbf{Acronym} \\
        \SetCell[r=7]{} Conferences & International Conference on Software Engineering & ICSE \\
        & Symposium on the Foundations of Software Engineering & FSE \\
        & International Conference on Evaluation and Assessment in Software Engineering & EASE \\
        & Mining Software Repositories & MSR \\
        & Empirical Software Engineering and Measurement & ESEM \\
        & International Conference on AI Engineering – Software Engineering for AI & CAIN \\
        & Conference on Software Analysis, Evolution, and Reengineering & SANER \\
        \SetCell[r=5]{} Journals & Empirical Software Engineering & EMSE \\
        & Information and Software Technology & IST \\
        & Journal of Systems and Software & JSS \\
        & Transactions on Software Engineering & TSE \\
        & Transactions on Software Engineering and Methodology & TOSEM \\
    \end{tblr}
\end{table}

\subsection{Study Selection}
\label{subsec:studySelection}
 
The limitation that motivated our decision not to rely on database search strings also extends to the study selection process. Because the use of LLMs to support research-related tasks is often embedded within methodological descriptions rather than explicitly stated in titles or abstracts, a conventional title–abstract screening approach is not sufficient in this context. Relying exclusively on this level of screening would therefore risk missing studies that satisfy IC1, thereby compromising the completeness of the study selection process. To mitigate this issue, we adopted a more comprehensive and iterative selection approach that extends beyond standard screening practices.

First, we manually downloaded all papers published in the selected journals and conferences. We then defined a set of keywords to be searched within the full text of each paper, aiming to filter the corpus and identify studies with the potential to meet our inclusion criteria. The following keyword set was used:

\begin{tcolorbox}[title=Search String, fonttitle=\small\bfseries, colframe=black, fontupper=\small]
\texttt{``Generative AI'' \textbf{OR} ``GenAI'' \textbf{OR} ``Large Language Model*'' \textbf{OR} ``Prompt'' \textbf{OR} ``LLM*'' \textbf{OR} ``Gemini'' \textbf{OR} ``Copilot'' \textbf{OR} ``ChatGPT'' \textbf{OR} ``LLaMA''}
\end{tcolorbox}

The terms included in the keyword set were selected to reflect the most commonly used terminology referring to LLMs. In addition to generic terms (e.g., ``LLM'', ``Generative AI''), we incorporated the names of representative and widely adopted models, such as ``LLaMA'', as these are sometimes used interchangeably with the broader term LLM. Including both general and model-specific terms increases the likelihood of capturing studies that refer to LLMs either conceptually or through specific implementations.

Although additional tools and models may also be considered, we expect that references to them would typically co-occur with general LLM-related terminology. Therefore, the defined search string was considered sufficiently comprehensive to capture relevant studies while maintaining a manageable scope.

To apply these keywords and refine the set of downloaded papers, we adopted an automated approach implemented in Python. Using PyMuPDF\footnote{\url{https://pymupdf.readthedocs.io/en/latest/}}, we searched the full text for the predefined keywords and highlighted each occurrence, excluding the references section. The code used in this process is available on GitHub\footnote{\url{https://github.com/victoriaogomes/pdf-keyword-search-api/releases/tag/v1.0.0}}.

Subsequently, two researchers independently assessed a subset of papers from one of the selected conferences (n = 11) to calibrate their interpretation of the inclusion and exclusion criteria. For each paper, all highlighted keyword occurrences were examined to determine whether the study satisfied the inclusion criteria. After completing this independent assessment, the researchers met to compare decisions, discuss discrepancies, and refine their shared understanding of the screening rules.

After calibration, the remaining papers were divided between the two researchers to ensure a balanced screening workload. Each researcher independently assessed their assigned subset using the predefined inclusion and exclusion criteria established during the calibration step.

\subsection{Data Extraction}
\label{subsec:dataExtraction}

To collect the data required to answer the research questions, we first designed a structured data extraction form. Two researchers then independently conducted a pilot extraction on three studies to validate the form and calibrate their understanding of the extraction procedure. The final version of the data extraction form is presented in Table~\ref{tab:data-collection-form}. The form was implemented in Google Sheets to provide a centralized and version-controlled repository for the extracted data. 

\begin{table}[ht]
\caption{Questions used for data collection from the selected studies}
\label{tab:data-collection-form}
\begin{tblr} {
    colspec = {Q[c, m, 0.3cm] Q[c, m, 1.9cm] X[j, m] Q[c, m, 0.6cm]},
    hlines,
    row{1} = {bg=black, fg=white, font=\bfseries},
    hline{15} = {1pt, solid},
    rows = {font=\footnotesize},
    rowsep = 1pt
}
    \textbf{Id} & \textbf{Data item} & \textbf{Description/Definition} & \textbf{RQ} \\
    1 & Bibliographic info & Title, authors, publication venue, and year of publication of the study & All \\
    2 & ESE task supported by LLM & Empirical Software Engineering (ESE) activity supported by the LLM & RQ1, RQ2, RQ3 \\
    3 & LLM model & Specific Large Language Model used & All \\
    4 & LLM version & Specific version of the LLM used & RQ5 \\
    5 & Model type & Type of LLM used (commercial/proprietary or open-source) & RQ5 \\
    6 & Model configuration & Configuration parameters controlling the LLM behavior (e.g., temperature, maximum token length) & RQ5 \\
    7 & Prompt development & Description of how the prompts used were designed or developed & RQ5 \\
    8 & Are the used prompts reported? & Whether the study reports the complete prompts used to instruct the LLM & RQ5 \\
    9 & Open LLM used as a baseline & Open-source LLM used as a baseline for performance comparison & RQ5 \\
    10 & Human validation for the LLM outputs & Whether and how human validation was performed on the outputs generated by the LLM & RQ5 \\
    11 & Dates experiments were conducted & Dates when the experiments involving the LLM were conducted & RQ5 \\
    12 & Reported benefits of using an LLM & Benefits reported regarding the use of LLMs to support empirical activities & RQ4 \\
    13 & Reported limitations of using an LLM & Limitations reported regarding the use of LLMs to support empirical activities & RQ4 \\
\end{tblr}
\end{table}

The data extraction process yielded two categories of information. The first category comprises descriptive data (items 1, 3-5, 8-9, 11 in Table~\ref{tab:data-collection-form}), which provide contextual information about the primary studies, support the analysis of trends and characteristics across them, and were straightforward to classify. The second category consists of qualitative data, corresponding to the remaining items, which capture detailed information on how LLMs were applied and discussed, enabling a deeper understanding of usage patterns and methodological practices. For these items, we collected verbatim excerpts from the primary studies to preserve contextual meaning and support accurate interpretation during analysis.
 
To answer RQ5, which investigates how studies report the experimental settings when using LLMs, the corresponding data extraction items were defined based on the reporting guidelines proposed by Wagner et al. \cite{wagner2025guidelines}. These guidelines specify the key information that should be documented when describing experiments involving LLMs, particularly to support transparent and reproducible LLM usage.

After calibration, the two reviewers independently extracted data from a subset of three papers, resulting in 39 individual coding decisions across the extracted fields. Disagreements occurred in 6 of these cases, yielding an overall agreement rate of 84.6\%. After resolving these disagreements through discussion, the remaining studies were divided between the two researchers for full data extraction. Each researcher then independently extracted data from their assigned subset using the predefined extraction form and the coding guidelines refined during calibration. Overall, uncertainties or ambiguities requiring discussion arose in 10 papers. These cases were jointly discussed and resolved through consensus to ensure consistency in the extracted data. One representative ambiguity involved studies in which LLMs were used both as part of the proposed solution and as support for conducting the empirical study. In such cases, it was necessary to distinguish between LLM use that constituted the research object or artifact under evaluation and LLM use that actually assisted an empirical software engineering research task, since only the latter was within the scope of this review.

\section{Overview of Primary Studies}
\label{sec:overview}

\begin{figure}[ht]
    \centering
    \includegraphics[width=\linewidth]{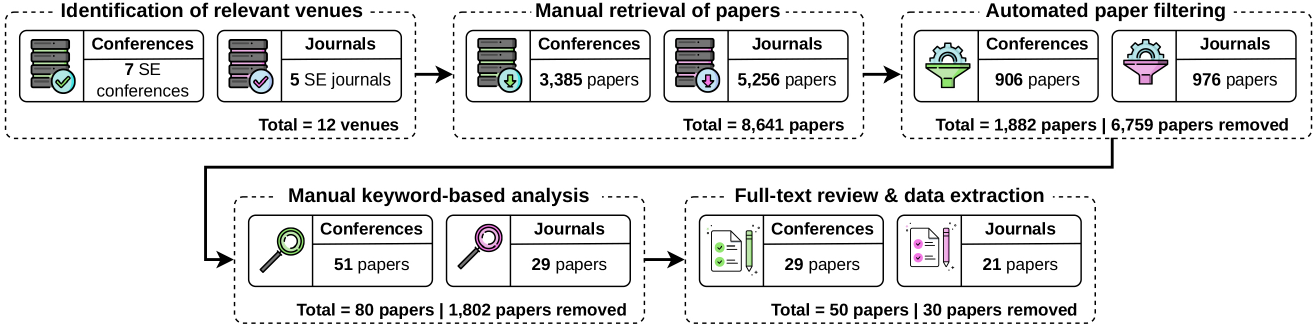}
    \caption{Study selection process.}
    \label{fig:study-selection-phase}
\end{figure}

Figure \ref{fig:study-selection-phase} presents the complete study selection process. We began by identifying relevant software engineering conferences and journals, resulting in a total of 12 selected venues. From these venues, we manually downloaded 8,641 papers published between 2020 and 2025. Next, we applied an automated filtering procedure to retain only papers containing at least one occurrence of the predefined keywords in the full text, which reduced the dataset to 1,882 papers. We then manually examined each keyword occurrence against the inclusion and exclusion criteria, further narrowing the set to 80 papers. Finally, we conducted a full-text review and data extraction. During this in-depth analysis, 30 additional papers were excluded for not fully satisfying the inclusion criteria, resulting in a final sample of 50 primary studies. To support transparency and traceability, the replication package \cite{replicationPackage} includes the papers considered at each stage of the study selection process, organized into separate tabs of the spreadsheet.

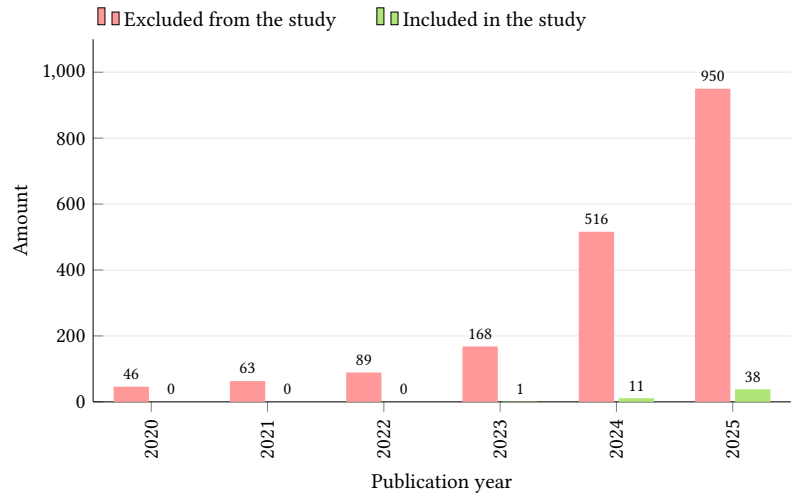
\begin{figure}[H]
    \centering   
    \definecolor{customGreen}{HTML}{ade576}

\begin{tikzpicture}
    \begin{axis}[
        ybar,
        height=7cm,
        width=12cm,
        bar width=15pt,
        ylabel={Amount},
        xlabel={Publication year},
        symbolic x coords={2020, 2021, 2022, 2023, 2024, 2025},
        xtick=data,
        nodes near coords,
        nodes near coords style={font=\footnotesize, /pgf/number format/fixed},
        ymin=0, ymax=1100,
        legend style={
            at={(0.36, 1.0)},
            anchor=south, 
            legend columns=-1, 
            draw=none,
            /tikz/every even column/.append style={column sep=15pt}
        },
        axis x line*=bottom,
        axis y line*=left,
        ymajorgrids=true,
        grid style={gray!20},
        xticklabel style={rotate=90, anchor=east}
    ]
        \addplot[fill=red!40, draw=none] coordinates {
            (2020, 46) (2021, 63) (2022, 89) (2023, 168) (2024, 516) (2025, 950)
        };
        
        \addplot[fill=customGreen, draw=none] coordinates {
            (2020, 0) (2021, 0) (2022, 0) (2023, 1) (2024, 11) (2025, 38)
        };
        
        \legend{Excluded from the study, Included in the study}
    \end{axis}
\end{tikzpicture}
    \caption{LLM papers accepted vs. rejected by year.}
    \label{fig:llm-per-year}
\end{figure}

We then proceed to analyze the demographics of the included papers. Figure \ref{fig:llm-per-year} presents the yearly distribution of the 1,882 papers retained after automated filtering (only papers containing at least one occurrence of the predefined keywords in the full text), distinguishing between those ultimately included in this review and those excluded. As expected, there is a clear upward trend in the number of papers mentioning LLM-related terms between 2020 and 2025. This growth becomes particularly pronounced after November 2022, following the release of ChatGPT. Notably, although the overall number of LLM-related publications increases substantially over time, only a relatively small fraction satisfies our inclusion criteria. 

The relatively small number of selected studies should be interpreted in relation to the specific boundary adopted in this review. Rather than synthesizing all software engineering studies involving LLMs, we focused on papers in which LLMs supported or were empirically examined within empirical software engineering research workflows. As a result, we excluded studies where LLMs were used primarily to support software engineering tools, techniques, or artifacts without contributing to an empirical research activity. For example, \cite{Lamed} was excluded because the LLM is used to generate annotations for memory leak detection, supporting a software analysis tool rather than an empirical research process. Borderline cases were assessed according to the same criterion. \cite{Adaptoring}, for instance, uses an LLM as a baseline when evaluating an adapter-generation tool, but the LLM serves as a comparison point for a software engineering technique rather than as support for an ESE task. Similarly, \cite{GreenRunner} refers to an LLM trained on scientific publications, but applies it to developer-facing component selection rather than to an empirical research workflow. The complete list of candidate papers, together with the applied inclusion and exclusion criteria and the corresponding selection decisions, is available in our replication package \cite{replicationPackage}.

From this perspective, the relatively small number of studies aligned with the scope of this review reinforces the interpretation that the systematic integration of LLMs into ESE workflows represents an emerging methodological shift rather than an established research practice. One possible explanation is that integrating LLMs into empirical research workflows raises methodological concerns related to transparency, reproducibility, and researcher oversight, which may slow their adoption despite the rapid growth of LLM-related research in software engineering.

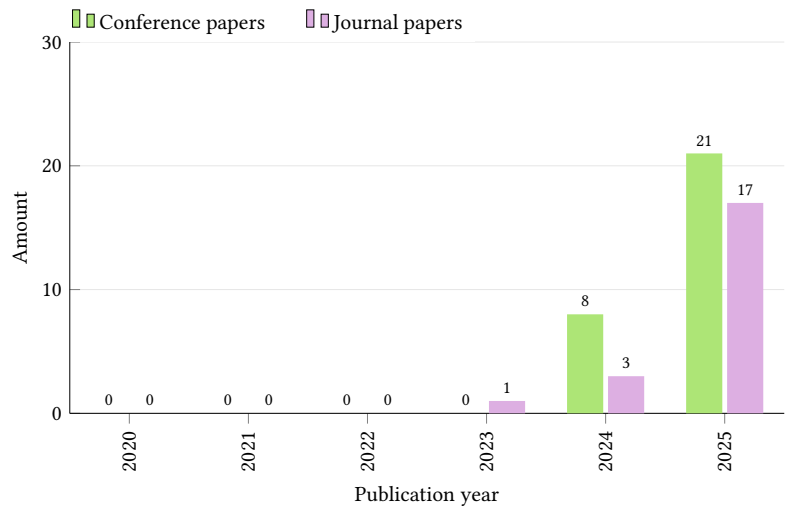
\begin{figure}[H]
    \centering   
    \definecolor{customGreen}{HTML}{ade576}
\definecolor{customPurple}{HTML}{ddafe2}

\begin{tikzpicture}
    \begin{axis}[
        ybar,
        height=7cm,
        width=12cm,
        bar width=15pt,
        ylabel={Amount},
        xlabel={Publication year},
        symbolic x coords={2020, 2021, 2022, 2023, 2024, 2025},
        xtick=data,
        nodes near coords,
        nodes near coords style={font=\footnotesize, /pgf/number format/fixed},
        ymin=0, ymax=30,
        legend style={
            at={(0.28, 1.0)},
            anchor=south, 
            legend columns=-1, 
            draw=none,
            /tikz/every even column/.append style={column sep=15pt}
        },
        axis x line*=bottom,
        axis y line*=left,
        ymajorgrids=true,
        grid style={gray!20},
        xticklabel style={rotate=90, anchor=east}
    ]
        \addplot[fill=customGreen, draw=none] coordinates {
            (2020, 0) (2021, 0) (2022, 0) (2023, 0) (2024, 8) (2025, 21)
        };
        
        \addplot[fill=customPurple, draw=none] coordinates {
            (2020, 0) (2021, 0) (2022, 0) (2023, 1) (2024, 3) (2025, 17)
        };
        
        \legend{Conference papers, Journal papers}
    \end{axis}
\end{tikzpicture}
    \caption{LLM papers accepted by year: Conferences vs Journals.}
    \label{fig:llm-per-venue-per-year}
\end{figure}

Figure \ref{fig:llm-per-venue-per-year} shows the distribution of selected papers across conferences and journals. The distribution aligns with the typical publication trajectory in software engineering, where new research topics predominantly appear in conferences first. This is evident in 2024 and 2025, where conference publications significantly outpace journal publications. While journal publications are also rising, suggesting the topic is quickly maturing to require the more extensive validation and reproducibility discussions typical of journal venues, conference venues remain the primary venue type for initial dissemination in this field.

\begin{figure}[H]
    \centering   
    \begin{tikzpicture}[font=\sffamily\footnotesize]
    \def\radius{3}
    \def\labeldist{4.2} 
    \def\lastangle{90}  
    
    \foreach \p/\c/\l/\n in {
        61.3/red!70/ChatGPT/38,
        1.6/orange!90/T5/1,
        6.5/orange!40/Claude/4,
        3.2/green!50/Not reported/2,
        1.6/gray!70/BERT/1,
        6.5/green!60!black!70/DeepSeek/4,
        1.6/red!40/Mixtral/1,
        3.2/cyan!60/Gemini/2,
        12.9/blue!40/Llama/8,
        1.6/blue!70/BingChat/1} {
        \pgfmathsetmacro{\startangle}{\lastangle}
        \pgfmathsetmacro{\endangle}{\lastangle - \p*3.6}
        \pgfmathsetmacro{\midangle}{(\startangle+\endangle)/2}

        \fill[\c, draw=white, line width=0.6pt] (0,0) -- (\startangle:\radius) arc (\startangle:\endangle:\radius) -- cycle;

        \pgfmathparse{cos(\midangle) < 0 ? "left" : "right"}
        \edef\side{\pgfmathresult}

        \draw[gray!60, thin] (\midangle:\radius) -- (\midangle:\labeldist) 
            node[\side, inner sep=2pt, black, align=\side] {
                \textbf{\l}\\
                \textcolor{gray}{\p\% (\n)}
            };

        \xdef\lastangle{\endangle}
    }
\end{tikzpicture}
    \caption{Distribution of the LLM models used across papers.}
    \label{fig:papers-per-llm-model}
\end{figure}
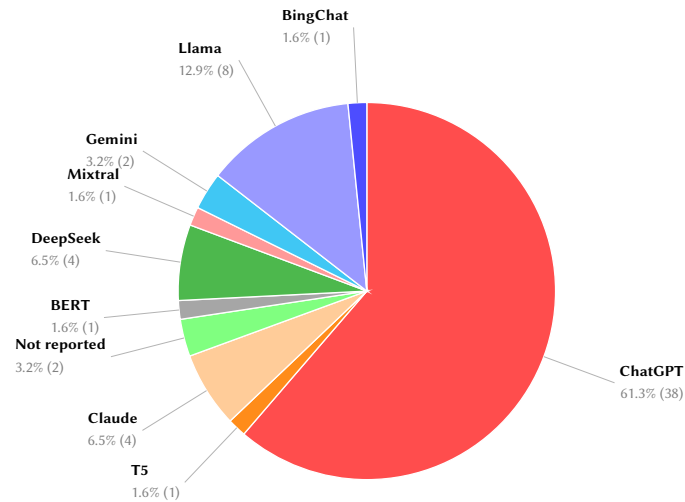

Finally, we examine the specific LLM models adopted in the selected studies. As shown in Figure \ref{fig:papers-per-llm-model}, the 50 analyzed papers account for 62 model usages. This number exceeds the total number of papers because several studies employed more than one LLM, either to compare model performance or to support different ESE tasks within the same study.

Overall, the LLM model distribution indicates a strong concentration around proprietary, general-purpose conversational models, with comparatively limited adoption of open or task-specific alternatives. ChatGPT is by far the most frequently used model, being applied 38 times. This dominant presence suggests that much of the empirical evidence on LLM-assisted ESE tasks is currently shaped by the capabilities, limitations, and design choices of a single model family. \textbf{While this concentration facilitates comparability across studies, it also introduces a potential ecosystem bias: conclusions about LLMs in general may, in practice, reflect the behavior of a specific commercial system}.

A noteworthy result is that only two studies did not report which LLM model was used. Transparent reporting of the exact model version, provider, and configuration is essential in ESE research involving AI-based systems, as these constitute fundamental experimental parameters. LLMs vary substantially in architecture, training data, alignment strategies, context window size, and update frequency. Without this information, reproducibility, comparability, and interpretability would be compromised. 
\section{Data Analysis}
\label{sec:dataAnalysis}

To answer \textbf{RQ1 - What empirical software engineering research tasks are supported by LLMs in LLM-assisted research studies?}, we first performed descriptive coding~\cite{saldana2025coding}, assigning short summarizing phrases to excerpts describing the ESE research task supported by LLMs. Then, to support a systematic analysis, we interpret the identified tasks through a lens that captures the role performed by the LLM within each task, such as semantic classifier, criteria-based screener, or evaluator. This abstraction allows us to reduce task-level granularity while preserving their functional characteristics, without shifting the focus away from the tasks themselves.

To answer \textbf{RQ2 - Which phases of the ESE research process are supported by LLMs?}, we analyze where LLM-supported tasks are situated within the empirical research lifecycle.

To identify the research phases, we first map the descriptive coding obtained for each task in RQ1 to their corresponding empirical research method using the taxonomy presented in Table~\ref{tab:empirical-study-types}. Because many studies combine multiple research strategies (e.g., mining software repositories and surveys), the research method is assigned according to the methodological component in which the LLM-supported task occurs. For example, if a study analyzes bug distributions in a dataset and uses an LLM to assist with bug classification, we classify it as mining software repositories because the LLM contributes to a step within that analysis. Conversely, if a study investigates the use of LLMs to perform title and abstract screening in systematic literature reviews, we classify the task as part of a systematic literature review, even if the paper in which it is presented is a controlled experiment.

\begin{table}[ht]
\caption{Empirical research methods and their steps.}
\label{tab:empirical-study-types}
\begin{tblr} {
    colspec = {Q[c, m, 2cm] X[j, m] Q[j, m, 3cm]},
    hlines,
    row{1} = {bg=black, fg=white, font=\bfseries},
    hline{8} = {1pt, solid},
    rows = {font=\footnotesize},
    rowsep = 1pt
}
    \textbf{Research method} & \textbf{Description} & \textbf{Steps} \\
    Case Study \cite{ExperimentationSEWohlin} & Empirical inquiry that investigates a contemporary phenomenon within its real-life context, especially when boundaries between phenomenon and context are not clear & Design, preparation for data collection, evidence collection, analysis of evidence, and reporting \\
    Controlled Experiment \cite{ExperimentationSEWohlin} & Strategy used to investigate cause-and-effect relationships by manipulating independent variables to observe their impact on dependent variables under controlled conditions & Scoping, planning, operation, analysis, and reporting \\
    Mining Software Repositories (MSR) \cite{Easterbrook2008} & Study of historical data found in software repositories to identify patterns, trends, and relationships in software development & Planning, data selection, extraction, pre-processing \& cleaning, analysis, and reporting \\
    Survey \cite{GuidelinesSurveySE} & Quantitative or qualitative strategy used to identify characteristics or opinions of a broad population by collecting data from a representative sample & Design, instrument construction, pilot testing, execution, data analysis, and reporting \\
    Systematic Literature Review (SLR) \cite{Kitchenham2007} & A secondary study used to identify, evaluate, and interpret all available research relevant to a specific research question & Planning, searching, screening, quality assessment, data extraction \& synthesis, and reporting \\
    Systematic Mapping Study (SMS) \cite{petersen2008systematic} & Secondary study designed to provide a high-level overview of a research area to identify clusters of evidence and research gaps & Planning, searching, screening, data extraction \& mapping, and reporting \\
\end{tblr}
\end{table}

After determining the research method, we map the coded tasks to the corresponding research steps associated with that method. The considered steps are presented in Table~\ref{tab:empirical-study-types}. For example, in a systematic literature review, an LLM used for title and abstract screening is mapped to the screening step, as it supports the inclusion/exclusion decision process within that methodological component.

To enable comparison across methods, we then abstract method-specific steps into five standardized research phases, following the general process of empirical software engineering research \cite{ExperimentationSEWohlin, Kitchenham2007}:
\begin{enumerate}
    \item \textbf{Planning \& Design:} Encompasses scoping, research question formulation, and selection of study protocols, instruments, and experimental designs to establish the theoretical and procedural framework.
    \item  \textbf{Data Collection:} Covers the active acquisition of evidence, including controlled experiments, surveys, repository mining, and literature searches for systematic reviews.
    \item \textbf{Data Processing:} Focuses on refining raw data through extraction, cleaning, pre-processing, pilot testing, and screening of artifacts or literature to ensure quality.
    \item \textbf{Analysis \& Synthesis:} Involves interpreting processed data to derive insights via statistical analysis, qualitative coding, or synthesis across primary studies.
    \item \textbf{Reporting:} Entails structured documentation, visualization of results, discussion of threats to validity, and dissemination to ensure reproducibility and communicability.
\end{enumerate}

Table~\ref{tab:standardized-phases} presents the mapping from method-specific steps to standardized phases. For example, the screening step in systematic literature reviews (SLRs) and the data extraction step in mining software repositories (MSR) are both mapped to the broader data processing phase. 

\begin{table}[ht]
\caption{Comprehensive Mapping of Method-Specific Steps to Standardized Research Phases.}
\label{tab:standardized-phases}
\begin{tblr}{
  colspec = {X[l,m] Q[l,m,2cm] Q[l,m,2cm] Q[l,m,1.9cm] Q[l,m,1.9cm] Q[l,m,1.6cm]},
  hlines,
  row{1} = {bg=black, fg=white, font=\bfseries, ht=1.3cm},
  hline{8} = {1pt, solid},
  rows = {font=\footnotesize},
  rowsep = 1pt
}
\diagbox[linewidth=1pt]{\makecell[l]{\textbf{Research} \\ \textbf{Method}}}{\makecell[l]{\textbf{Standardized} \\ \textbf{phase}}} & Planning \& Design & Data Collection & Data Processing & Analysis \& Synthesis & Reporting \\
Case Study & Design, Preparation for data collection & Evidence collection & --- & Analysis of evidence & Reporting \\
Controlled Experiment & Scoping, Planning & Operation & --- & Analysis & Reporting \\
Mining Software Repositories & Planning & Data selection & Data extraction, Pre-processing \& cleaning & Analysis & Reporting \\
Survey & Design, Instrument construction & Execution & Pilot testing & Data analysis & Reporting \\
Systematic Literature Review & Planning & Searching & Screening, Quality assessment, Data Extraction & Data synthesis & Reporting \\
Systematic Mapping Study & Planning & Searching & Data extraction & Mapping & Reporting \\
\end{tblr}
\end{table}

To answer \textbf{RQ3 - How are LLMs integrated into empirical software engineering research workflows?}, we analyze how LLMs contribute to the execution of empirical tasks, focusing on their modes of involvement and interaction with human researchers.

The contribution type is determined through a joint analysis of the roles identified in RQ1 via descriptive coding and the corresponding excerpts describing how the LLM supports each task. This enables us to characterize how LLMs are integrated into workflows, whether they replace, support, or collaborate with human effort. Based on this analysis, we classify LLM involvement into four categories:

\begin{itemize}
    \item \textbf{Augmentation} refers to collaborative scenarios in which the LLM enhances human performance by generating intermediate outputs, reformulating artifacts, or accelerating parts of the task while still requiring active human involvement;
    \item \textbf{Automation} refers to cases in which the LLM performs a task with minimal or no human intervention, effectively replacing manual effort;
    \item \textbf{Decision support} describes situations where the LLM provides suggestions, recommendations, or analyses that assist researchers in decision-making while leaving the final judgment to humans;
    \item \textbf{Evaluation} refers to cases where the LLM is used to assess, review, or validate artifacts or results produced during the research process. 
\end{itemize}

This categorization captures different levels of human-AI collaboration, as discussed in the literature on automation and decision-support systems \cite{parasuraman2000model, shneiderman2020human}. Rather than being determined solely by the task itself, the contribution type depends on how the LLM is integrated into the research workflow; therefore, the same task may correspond to different contribution types across studies.

To address \textbf{RQ4 - What benefits and limitations are reported when using LLMs to support empirical software engineering research tasks?}, we conducted a thematic analysis on the excerpts extracted from selected studies, focusing on the reported benefits and limitations of LLM usage. The analysis followed established guidelines in software engineering research \cite{ThematicAnalysisSE} and was structured as a hierarchical abstraction process, progressing from \textit{quote} to \textit{code}, then to \textit{lower-level themes}, and finally to \textit{higher-level themes}. This systematic approach ensured both analytical rigor and traceability across all levels of interpretation.

Finally, to answer \textbf{RQ5 - To what extent do studies proposing LLM-assisted empirical software engineering research report the information required to enable reproducible use of LLMs?}, we conducted an analysis based on the six guidelines proposed by Wagner et al. \cite{wagner2025guidelines}, which aim to support the transparency and reproducibility of LLM usage, an aspect that is central to this research question.

Guided by these recommendations, we collected the relevant information corresponding to items 2-11 in the data collection form (see Table~\ref{tab:data-collection-form}) by analyzing both the primary studies and their replication packages. We then formulated nine assessment questions and answered each of them with ``Yes" or ``No" for every LLM-assisted task reported in the analyzed papers. The questions are presented in Table \ref{tab:assessment-questions}. Lastly, we analyzed these data to assess how studies report and document LLM-assisted tasks.

\begin{table}[ht]
\caption{Assessment Questions for Evaluating Transparency and Reproducibility of LLM-Assisted Tasks.}
\label{tab:assessment-questions}
\begin{tblr} {
    colspec = {Q[c, m, 0.5cm] X[j, m]},
    hlines,
    row{1} = {bg=black, fg=white, font=\bfseries},
    hline{11} = {1pt, solid},
    rows = {font=\footnotesize},
    rowsep = 1pt
}
    \textbf{Id} & \textbf{Question} \\
    Q1 & Does the study declare the use of an LLM and describe its role? \\
    Q2 & Does the study report the model version? \\
    Q3 & Does the study report the date when the experiments were conducted? \\
    Q4 & Does the study report the model configuration? \\
    Q5 & Does the study report the prompts used? \\
    Q6 & Does the study report how the prompts were developed? \\
    Q7 & Does the study use an open LLM as a baseline? \\
    Q8 & Does the study use human validation for the LLM outputs? \\
    Q9 & Does the study report how the human validation of the LLM outputs was conducted? \\
\end{tblr}
\end{table}

\section{Analysis of the Results}
\label{sec:results}

This section presents the findings of our systematic literature review, structured around the five research questions defined in Section~\ref{sec:methodology}. Building on the dataset of 50 primary studies described in Section~\ref{sec:overview}, we analyze how LLMs are used to support empirical software engineering research, examining both their functional roles and their methodological implications. Specifically, we first characterize the empirical research tasks supported by LLMs (RQ1), followed by an analysis of how these tasks are distributed across the phases of the empirical research lifecycle (RQ2). We then investigate how LLMs are integrated into research workflows by analyzing their modes of involvement and interaction with human researchers (RQ3). Next, we synthesize the benefits and limitations reported in the literature (RQ4), and finally assess the extent to which current studies provide the information necessary to ensure transparency and reproducibility in LLM-assisted research (RQ5).

\subsection{RQ1 - What empirical software engineering research tasks are supported by LLMs in LLM-assisted research studies?}
To answer this research question, we applied descriptive coding to excerpts from the selected studies that described the ESE tasks supported by LLMs. This process resulted in 69 distinct tasks. Table~\ref{tab:example-tasks} presents 10 representative examples of these tasks, while the complete list is available in the replication package \cite{replicationPackage}.

\begin{table}[ht]
\caption{Examples of identified LLM-supported empirical software engineering research tasks.}
\label{tab:example-tasks}
\begin{tblr}[
] {
    colspec = {Q[c, m, 2.8cm] X[j, m] Q[c, m, 1.3cm]},
    hlines,
    row{1} = {bg=black, fg=white, font=\bfseries},
    hline{12} = {1pt, solid},
    rows = {font=\footnotesize},
    rowsep = 1pt
}
    \textbf{LLM task} & \textbf{Description} & \textbf{Paper ID} \\
    Evaluation of code comments & Uses LLMs as evaluators to assess the quality of generated Java method comments, following established evaluation prompts from prior work. & A63 \\
    Title-abstract screening of papers & Applies LLMs to automatically screen paper titles and abstracts based on predefined inclusion and exclusion criteria, replicating a traditional manual screening process. & A105 \\
    Generation of values to fill placeholders & Uses LLMs to generate synthetic values for placeholders in developer prompts, ensuring consistent and sequential filling when multiple fields are present. & A142 \\
    Commit message filtering & Uses an LLM to refine regex-filtered commit messages by semantically classifying them against the ``2024 CWE Top 25" list to identify security-related commits. & A152 \\
    Bug-report classification & Uses instruction-tuned LLMs to first identify bug reports and then classify them as responsive or non-responsive across npm packages. & A180 \\
    Evaluate the quality of Jupyter Notebooks & Uses an LLM as an automated evaluator to assess the quality of Jupyter Notebooks generated from sketches, assigning scores based on a structured evaluation prompt. & A445 \\
    Generation of user stories & Uses LLMs to generate realistic user stories based on a specified machine learning task or technique within a given domain. & A671 \\
    Scoring of pairs of code and code review & Uses LLMs to score code–review pairs on a 10-point scale, retaining only high-quality reviews based on a predefined threshold. & A767 \\
    Classification of review comments & Uses LLMs to classify code review comments as accurate, partially accurate, or not accurate, and to assess alignment with human judgments. & A767 \\
    Extraction of vulnerability information & Uses LLMs to extract vulnerability-related information from web pages by summarizing content and performing binary (Yes/No) classification for predefined aspects. & A776 \\    
\end{tblr}
\end{table}

To synthesize the identified tasks while preserving their functional characteristics, we abstracted them into higher-level functional roles (e.g., classification, extraction, or evaluation). These roles do not replace tasks as the unit of analysis, but rather provide an analytical lens to group semantically similar tasks. Table~\ref{tab:identified-llm-roles} presents the identified roles, along with their descriptions and the number of occurrences. Consistent with our analysis procedure, a role refers to the specific function performed by the LLM within a task (e.g., semantic classifier or criteria-based screener).

The analysis reveals a clear concentration of LLM applications in tasks centered around classification, filtering, and evaluation. The most frequent role, \textit{semantic classifier} (17 tasks), demonstrates that LLMs are primarily leveraged to assign discrete labels to inputs based on semantic content, reflecting their strength in understanding and categorizing textual or code artifacts. The high occurrence of \textit{criteria-based screener} (14 tasks) complements this, indicating that LLMs are also trusted to enforce explicit decision rules, such as inclusion/exclusion in systematic reviews or relevance filtering in large datasets. Together, these two roles account for nearly half of the identified tasks, suggesting that current LLM adoption is largely oriented toward structured decision-making activities.

Beyond classification and screening, LLMs are increasingly applied in evaluative and generative functions. The \textit{evaluator} role (11 tasks) shows that LLMs can provide qualitative or quantitative assessments, such as scoring commit messages or code artifacts, bridging the gap between automated analysis and human judgment. Generative roles like \textit{synthetic data generator} (8 tasks) and \textit{content transformer} (5 tasks) highlight that LLMs are capable of augmenting datasets or reformatting content while preserving semantic integrity, enabling experimentation, simulation, and data enrichment in ways traditional tools cannot.

\begin{table}[ht]
\caption{Higher-level functional roles derived from LLM-supported tasks.}
\label{tab:identified-llm-roles}
\begin{tblr} {
    colspec = {Q[c, m, 1.5cm] X[j, m] Q[c, m, 1.5cm] Q[c, m, 2.5cm]},
    hlines,
    row{1} = {bg=black, fg=white, font=\bfseries},
    hline{11} = {1pt, solid},
    rows = {font=\footnotesize},
    rowsep = 1pt
}
    \textbf{LLM role} & \textbf{Description} & \textbf{N. of tasks} & \textbf{Paper Ids}\\
    Semantic Classifier & The LLM assigns inputs to predefined categories based on their semantic content, typically producing discrete labels (e.g., mapping commits to maintenance activities) & 17 & A46, A64, A96, A152, A154, A180, A447, A806, A925, A936, A939, A966, A1106, A1354, A1370, A1455, A1660 \\
    Criteria-Based Screener & The LLM filters items by applying explicit inclusion and exclusion criteria, producing binary or rule-based relevance decisions (e.g., title-abstract screening for SLRs) & 14 & A64, A105, A125, A136, A152, A154, A171, A180, A447, A806, A877, A1109, A1132, A1435 \\
    Evaluator (LLM-as-a-Judge) & The LLM performs qualitative or quantitative assessment of artifacts by assigning scores, rankings, or judgments based on predefined or implicit evaluation criteria (e.g., scoring commit messages on a 10-point scale) & 11 & A46, A63, A142, A161, A440, A445, A767, A779, A1660 \\
    Synthetic Data Generator & The LLM generates novel artifacts independent of the input data to simulate scenarios, augment datasets, or create hypothetical instances (e.g., creating software personas) & 8 & A134, A142, A227, A671, A787, A1004, A1508 \\
    Content Transformer & The LLM modifies the form, style, structure, or granularity of an existing artifact while preserving its semantic content and factual integrity (e.g., abstract simplification) & 5 & A46, A105, A926, A1106, A1673 \\
    Pattern \& Relational Analyst & The LLM identifies patterns, relationships, or latent structures across multiple data points, enabling tasks such as clustering, thematic analysis, and comparative reasoning (e.g., performing inductive thematic analysis) & 5 & A117, A1062, A1060, A1096, A1315 \\
    Structured Data Extractor & The LLM extracts specific entities, attributes, or fields from unstructured data and converts them into structured representations (e.g., extracting build error codes) & 5 & A140, A523, A751, A776, A1106 \\
    Logic \& Consistency Auditor & The LLM verifies logical coherence and consistency within or between artifacts, detecting contradictions, violations, or misalignments (e.g., detecting contradictions in prompt instructions) & 3 & A64, A806, A1870 \\
    Semantic Mapper & The LLM bridges distinct domains or abstraction levels, translating concepts or elements from one domain to another (e.g., mapping high-level financial tokenomics descriptions to specific low-level code variable names) & 1 & A1435\\
\end{tblr}
\end{table}

Overall, the distribution of roles suggests that LLMs are widely applied to tasks that are repetitive, well-defined, and semantically constrained, while their potential for complex reasoning, cross-domain abstraction, and advanced data analysis is still underutilized. This gap points to promising directions for future research, particularly in leveraging LLMs for tasks that require integrative understanding across heterogeneous datasets, automated auditing of logical consistency, and creative problem generation in software engineering contexts.

\begin{tcolorbox}[title=Summary of RQ1, fonttitle=\small\bfseries, colframe=black, fontupper=\small]
\begin{itemize}[leftmargin=*, itemsep=2pt, topsep=2pt]
\item LLM usage is concentrated in structured tasks, particularly classification, screening, and evaluation.
\item These tasks are semantically constrained and decision-oriented, leveraging core LLM strengths.
\item More complex roles, such as cross-domain reasoning and integrative analysis, remain underexplored.
\item Current usage patterns suggest a focus on efficiency rather than advanced analytical capabilities.
\end{itemize}
\end{tcolorbox}

\subsection{RQ2 - Which phases of the ESE research process are supported by LLMs in LLM-assisted studies?}
For visualization purposes in this RQ, we grouped research methods that share methodological proximity and overlapping analytical workflows. Accordingly, Systematic Literature Reviews (SLRs) were combined with Systematic Mapping Studies (SMSs), while Controlled Experiments were combined with Case Studies.

\begin{figure}[ht]
    \centering   
    \definecolor{customGreen}{HTML}{ade576}

\begin{tikzpicture}
    \begin{axis}[
        ybar,
        height=7cm,
        width=12cm,
        bar width=15pt,
        ylabel={Task count},
        xlabel={Research method},
        symbolic x coords={Controlled Exp. + Case Study, MSR, Survey, SLR + SMS},
        xtick=data,
        nodes near coords,
        nodes near coords style={font=\footnotesize, /pgf/number format/fixed},
        ymin=0, ymax=40,
        legend style={
            at={(0.07, 1.0)},
            anchor=south, 
            legend columns=-1, 
            draw=none,
            /tikz/every even column/.append style={column sep=15pt}
        },
        axis x line*=bottom,
        axis y line*=left,
        ymajorgrids=true,
        grid style={gray!20},
        xticklabel style={align=center, text width=4.3cm, yshift=-2pt}
    ]
        \addplot[fill=customGreen, draw=none] coordinates {
            (Controlled Exp. + Case Study, 21) (MSR, 34) (Survey, 3) (SLR + SMS, 11)
        };
        
        \legend{Tasks}
    \end{axis}
\end{tikzpicture}
    \caption{Number of LLM-assisted tasks by research method.}
    \label{fig:task-by-research-method}
\end{figure}
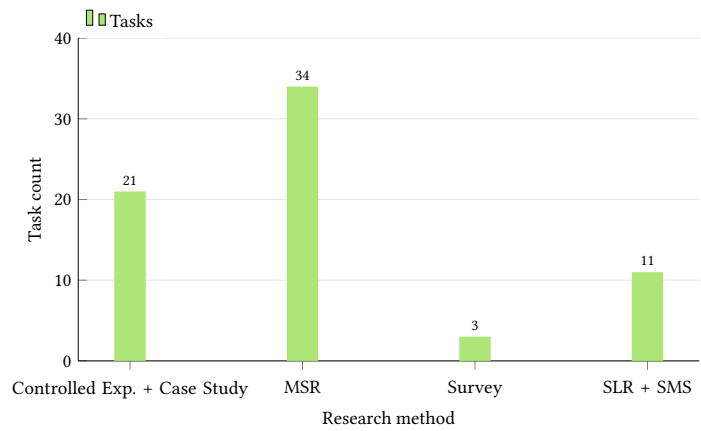

Figure~\ref{fig:task-by-research-method} illustrates how the 69 LLM-supported tasks are distributed across research methods. Mining software repositories (MSR) and controlled experiments and case studies dominate the landscape, with 34 and 21 tasks, respectively. This reflects the strengths of LLMs in processing large-scale, structured software data and supporting experimental workflows, where automation can reduce repetitive work and accelerate analysis. In contrast, methods that rely heavily on human judgment or qualitative reasoning, such as surveys, show relatively few tasks, indicating that LLM adoption is still limited in these areas. Secondary studies (systematic mapping studies (SMS) and SLRs) occupy an intermediate position, reflecting the emerging use of LLMs to assist data extraction and synthesis in secondary studies.

\begin{figure}[H]
    \centering
    \includegraphics[width=0.61\linewidth]{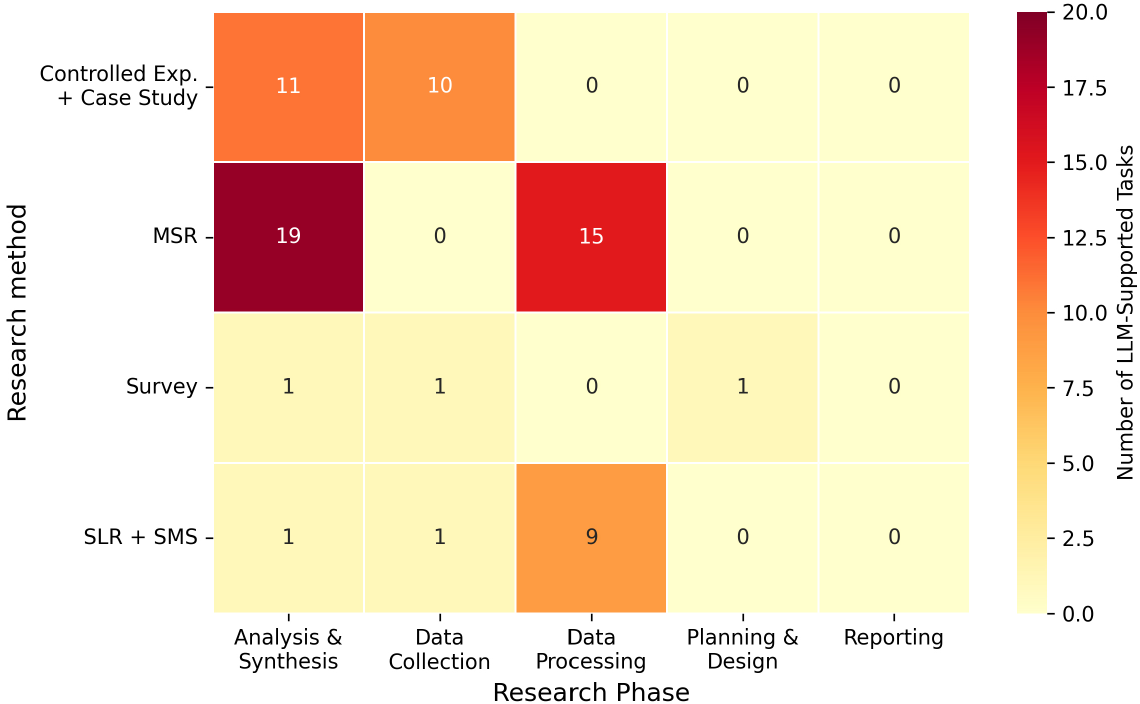}
    \caption{Heatmap illustrating the distribution of the identified LLM-assisted tasks across research methods and stages.}
    \label{fig:heatmap-llm-task-by-phase}
\end{figure}

To further understand how LLMs support specific phases across research methods, we identified the step associated with each task and mapped it to a standardized research phase (Table~\ref{tab:standardized-phases}). The heatmap in Figure ~\ref{fig:heatmap-llm-task-by-phase} shows that tasks appear across different phases within each research method. For example, in MSR, LLMs are applied both in \textit{data processing} and \textit{analysis \& synthesis}, while in controlled experiments and case studies they support \textit{data collection} as well as \textit{analysis \& synthesis}. Even in less frequent methods such as SLR/SMS and surveys, LLMs appear in multiple phases, including \textit{planning \& design}, \textit{data collection}, \textit{data processing}, and \textit{analysis \& synthesis}. This spread indicates that LLMs are being empirically explored across a range of research phases, rather than being confined to a single part of the workflow.

Lastly, although the heatmap indicates no reported use of LLMs in the reporting phase, a few studies mention LLM assistance in this context. These papers, however, were not included in our analysis because they provide insufficient detail on how the LLM was actually used. Typically, they merely note GenAI utilization in a small section of the paper without even specifying parts of the reporting process where the LLM contributed.

\begin{tcolorbox}[title=Summary of RQ2, fonttitle=\small\bfseries, colframe=black, fontupper=\small]
\begin{itemize}[leftmargin=*, itemsep=2pt, topsep=2pt]
\item LLM-supported tasks span multiple phases of the empirical research lifecycle.
\item Usage is concentrated in data processing and analysis phases.
\item Early-stage (planning and design) and late-stage (reporting) phases show limited adoption.
\item This indicates an uneven integration of LLMs across the full empirical workflow.
\end{itemize}
\end{tcolorbox}

\subsection{RQ3 - How are LLMs integrated into empirical software engineering research workflows?}

\begin{figure}[H]
    \centering
    \definecolor{customGreen}{HTML}{ade576}
\definecolor{customPurple}{HTML}{ddafe2}


\begin{tikzpicture}[font=\sffamily\footnotesize]
    \def\radius{3}
    \def\labeldist{4.2} 
    \def\lastangle{90}  
    
    \foreach \p/\c/\l/\n in {
        60.9/customGreen/Automation/42,
        18.8/customPurple/Augmentation/13,
        18.8/blue!40/Evaluation/13,
        1.5/orange!40/Decision Support/1} {
        \pgfmathsetmacro{\startangle}{\lastangle}
        \pgfmathsetmacro{\endangle}{\lastangle - \p*3.6}
        \pgfmathsetmacro{\midangle}{(\startangle+\endangle)/2}

        \fill[\c, draw=white, line width=0.6pt] (0,0) -- (\startangle:\radius) arc (\startangle:\endangle:\radius) -- cycle;

        \pgfmathparse{cos(\midangle) < 0 ? "left" : "right"}
        \edef\side{\pgfmathresult}

        \draw[gray!60, thin] (\midangle:\radius) -- (\midangle:\labeldist) 
            node[\side, inner sep=2pt, black, align=\side] {
                \textbf{\l}\\
                \textcolor{gray}{\p\% (\n)}
            };

        \xdef\lastangle{\endangle}
    }
\end{tikzpicture}
    \caption{Distribution of tasks according to LLM involvement categories.}
    \label{fig:task-by-llm-involvement}
\end{figure}
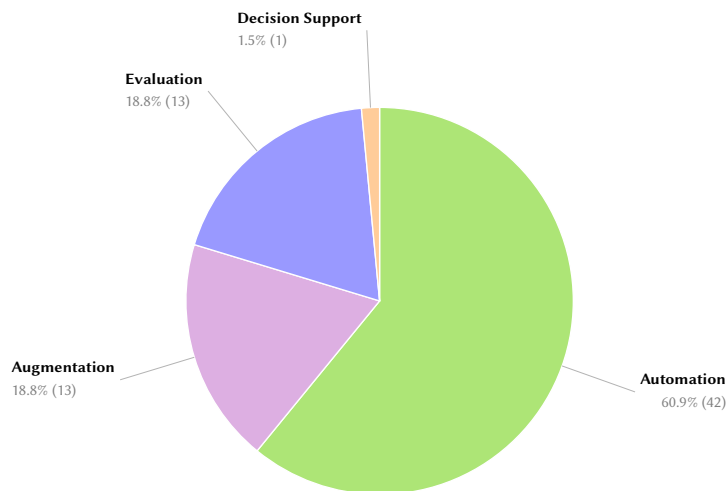

Building on the LLM roles identified in RQ1, we classify the nature of the LLM role into four categories: \textit{augmentation}, \textit{automation}, \textit{decision support}, and \textit{evaluation}. Figure~\ref{fig:task-by-llm-involvement} summarizes the distribution of tasks across these categories, providing a complementary view of how LLMs are integrated into empirical workflows.

The category \textit{automation} dominates, accounting for 42 tasks. This indicates that LLMs are frequently used to perform tasks with minimal or no human intervention, suggesting a tendency toward replacing manual effort in parts of the empirical workflow. Both \textit{augmentation} and \textit{evaluation} are represented in 13 tasks each, reflecting that LLMs are also leveraged to enhance human work, by generating suggestions, refining content, or assessing artifacts, without fully replacing human judgment. In contrast, \textit{decision support} appears in only one task, indicating that LLMs are rarely used to assist human decision-making. This contrasts with their frequent use in structured decision-making roles (e.g., classification and screening), where the LLM itself produces decisions rather than supporting human judgment.

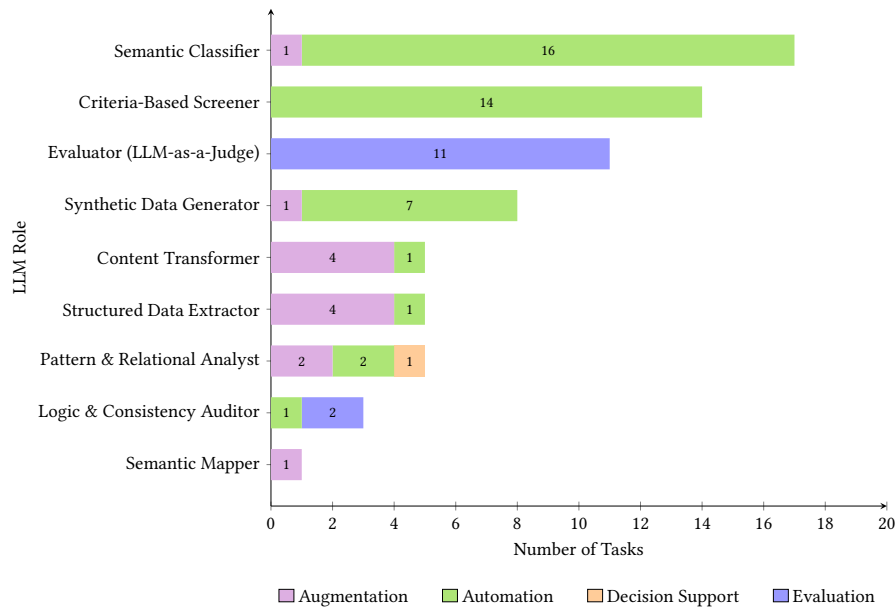
\begin{figure}[ht]
    \centering
    \definecolor{customPurple}{HTML}{ddafe2}
\definecolor{customGreen}{HTML}{ade576}

\begin{tikzpicture}
    \begin{axis}[
        xbar stacked,
        width=12cm,
        height=10cm,
        xmin=0,
        xmax=20,
        bar width=15pt,
        xlabel={Number of Tasks},
        ylabel={LLM Role},
        axis lines=left,
        symbolic y coords={
            Semantic Mapper,
            Logic \& Consistency Auditor,
            Pattern \& Relational Analyst,
            Structured Data Extractor,
            Content Transformer,
            Synthetic Data Generator,
            Evaluator (LLM-as-a-Judge),
            Criteria-Based Screener,
            Semantic Classifier
        },
        ytick=data,
        enlarge y limits=0.1,
        legend style={
            at={(0.5,-0.15)},
            anchor=north,
            legend columns=4,
            draw=none,
            /tikz/every even column/.append style={column sep=0.5cm}
        },
        every axis plot/.append style={fill, draw=none},
        nodes near coords, 
        nodes near coords align={center}, 
        every node near coord/.append style={font=\footnotesize} 
    ]
        
        \addplot[fill=customPurple] coordinates {
            (4,Content Transformer)
            (0,Criteria-Based Screener)
            (0,{Evaluator (LLM-as-a-Judge)})
            (0,Logic \& Consistency Auditor)
            (2,Pattern \& Relational Analyst)
            (1,Semantic Classifier)
            (1,Semantic Mapper)
            (4,Structured Data Extractor)
            (1,Synthetic Data Generator)
        };

        \addplot[fill=customGreen] coordinates {
            (1,Content Transformer)
            (14,Criteria-Based Screener)
            (0,{Evaluator (LLM-as-a-Judge)})
            (1,Logic \& Consistency Auditor)
            (2,Pattern \& Relational Analyst)
            (16,Semantic Classifier)
            (0,Semantic Mapper)
            (1,Structured Data Extractor)
            (7,Synthetic Data Generator)
        };

        \addplot[fill=orange!40] coordinates {
            (0,Content Transformer)
            (0,Criteria-Based Screener)
            (0,{Evaluator (LLM-as-a-Judge)})
            (0,Logic \& Consistency Auditor)
            (1,Pattern \& Relational Analyst)
            (0,Semantic Classifier)
            (0,Semantic Mapper)
            (0,Structured Data Extractor)
            (0,Synthetic Data Generator)
        };

        \addplot[fill=blue!40] coordinates {
            (0,Content Transformer)
            (0,Criteria-Based Screener)
            (11,{Evaluator (LLM-as-a-Judge)})
            (2,Logic \& Consistency Auditor)
            (0,Pattern \& Relational Analyst)
            (0,Semantic Classifier)
            (0,Semantic Mapper)
            (0,Structured Data Extractor)
            (0,Synthetic Data Generator)
        };

        \legend{Augmentation, Automation, Decision Support, Evaluation}
        
        \end{axis}
    \end{tikzpicture}
    \caption{Distribution of LLM Involvement by Role.}
    \label{fig:llm-involvement-by-role-and-task}
\end{figure}

Figure~\ref{fig:llm-involvement-by-role-and-task} illustrates the joint distribution of LLM roles and their nature, revealing a strong alignment between functional specialization and the degree of human intervention. The most prominent pattern is the concentration of automation-oriented tasks in roles that encode explicit decision rules, particularly \textit{semantic classifier} and \textit{criteria-based screener}. These roles account for the largest share of tasks and are predominantly associated with full \textit{automation}, indicating that LLMs are mainly used as end-to-end mechanisms for classification and filtering rather than as supporting components within human-driven workflows. 

A similar concentration is observed in evaluation-related roles. The \textit{evaluator} role is almost exclusively mapped to the \textit{evaluation} category, with a smaller contribution from \textit{logic \& consistency auditor}. This pattern suggests that evaluative functions are typically implemented as standalone processes, where the LLM directly produces judgments, scores, or consistency checks, with limited integration into broader interactive decision workflows.

In contrast, \textit{augmentation} roles such as \textit{content transformer} and \textit{structured data extractor} exhibit a more distributed behavior across categories. These roles appear in both \textit{augmentation} and \textit{automation} settings, suggesting that their use is not tied to a single mode of operation and may vary depending on the level of human oversight. This pattern indicates that transformation and extraction tasks can act as flexible building blocks, either supporting human analysis or being integrated into automated pipelines.

\textit{Decision support} remains marginal across the cross-analysis, with only a single task observed in this category. Notably, even analytically oriented roles, such as \textit{pattern \& relational analyst}, rarely manifest in a decision-support configuration, reinforcing the limited adoption of LLMs as tools for assisting rather than performing decisions.

Overall, this highlights a clear dichotomy in current empirical usage: LLMs are primarily deployed either in fully automated decision-making pipelines or in isolated evaluation tasks, while human-centered decision support remains largely underexplored, highlighting an opportunity for more collaborative human-AI interaction.

\begin{tcolorbox}[title=Summary of RQ3, fonttitle=\small\bfseries, colframe=black, fontupper=\small]
\begin{itemize}[leftmargin=*, itemsep=2pt, topsep=2pt]
\item LLMs are predominantly used for automation, often replacing manual effort.
\item Classification and screening tasks are frequently fully automated.
\item Augmentation and evaluation roles are present but less dominant.
\item Decision-support usage is rare, indicating limited human-AI collaboration.
\end{itemize}
\end{tcolorbox}

\subsection{RQ4 - What benefits and limitations are reported when using LLMs to support empirical software engineering research tasks?}
To address RQ4, we conducted two complementary thematic analyses: one focused on identifying the benefits reported in the literature and another aimed at synthesizing the reported limitations. 

\begin{figure}[ht]
    \centering
    \includegraphics[width=\linewidth]{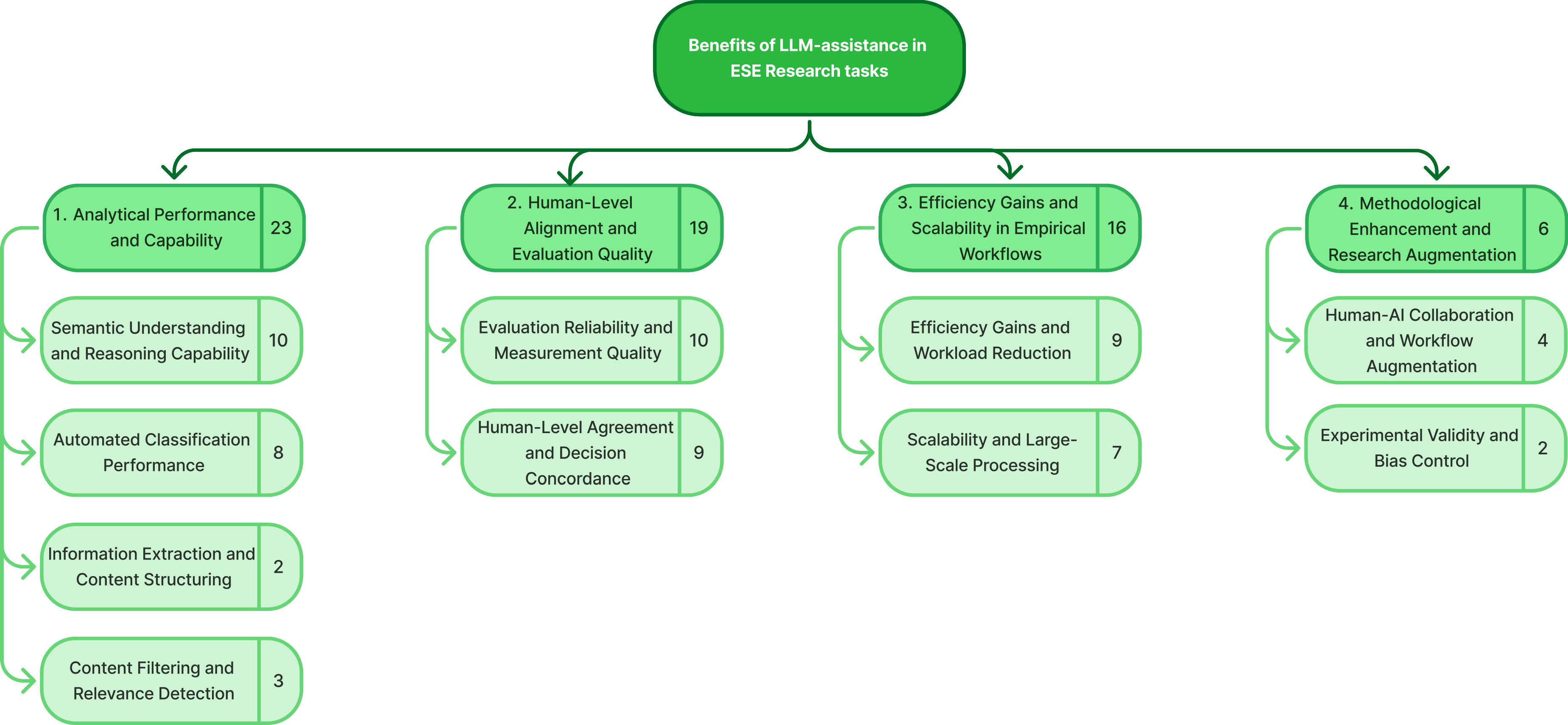}
    \caption{Thematic map of LLM benefits in software engineering research.}
    \label{fig:llm-benefits}
\end{figure}

\textbf{With respect to benefits}, the identified advantages of LLMs can be organized into four higher-level themes, as illustrated in Figure \ref{fig:llm-benefits}. 

These themes were supported by a total of 64 coded instances extracted from 33 of the 50 selected papers; the remaining 17 papers did not report any benefits. The four higher-level themes are defined as follows:
\begin{enumerate}
    \item \textit{Analytical Performance and Capability} captures the analytical value attributed to LLMs, particularly their ability to support cognitively demanding tasks such as understanding, structuring, filtering, and classifying research data;
    \item \textit{Human-Level Alignment and Evaluation Quality} emphasizes the extent to which LLM outputs approximate human judgment and, in doing so, contribute to the reliability, quality, and trustworthiness of evaluation processes;
    \item \textit{Efficiency Gains and Scalability in Empirical Workflows} highlights the role of LLMs in reducing manual effort and enabling empirical tasks to be conducted more quickly and at greater scale;
    \item \textit{Methodological Enhancement and Research Augmentation} encompasses benefits related to the strengthening of research practice, especially through human-AI collaboration and support for more rigorous empirical procedures.
\end{enumerate}

The lower-level themes associated with each higher-level category, together with their definitions, are presented in Table \ref{tab:llm_benefit_themes}.

\begin{longtblr}[
    caption={Definitions of the lower-level themes and their relationship to the higher-level benefit themes.},
    label={tab:llm_benefit_themes}
]{
    colspec = {Q[c, m, 2.45cm] Q[j, m, 2.95cm] X[j, m] Q[l, m, 2.2cm]},
    hlines,
    hline{12} = {1pt, solid},
    row{1} = {bg=black, fg=white, font=\bfseries},
    rows = {font=\footnotesize},
    rowsep = 1pt
}
    \textbf{Higher-level theme} & \textbf{Lower-level themes} & \textbf{Description} & \textbf{Ids}  \\
    \SetCell[r=4]{c} Analytical Performance and Capability & Semantic Understanding and Reasoning Capability & Reflects the value of LLMs in interpreting meaning, capturing context, and supporting reasoning over textual or qualitative material in ways that enhance empirical analysis & A447, A776, A1062, A925, A1096, A1106, A1109, A1673 \\ 
    & Automated Classification Performance & Captures the benefit of using LLMs to automatically categorize, label, or assign instances to predefined analytical classes with satisfactory performance & A46, A180, A440, A939, A966, A1109, A1315, A1870 \\
    & Information Extraction and Content Structuring & Highlights the contribution of LLMs in extracting relevant information from unstructured sources and organizing it into more structured and usable representations for analysis & A117  \\
    & Content Filtering and Relevance Detection & Represents the advantage of using LLMs to identify relevant content, remove irrelevant material, and support the selection or prioritization of information for empirical tasks & A877, A1132, A1455\\
    \SetCell[r=2]{c} Efficiency Gains and Scalability in Empirical Workflows & Efficiency Gains and Workload Reduction & Reflects the benefit of reducing time, effort, and repetitive manual work when LLMs are used to support empirical research activities & A105, A136, A180, A445, A936, A966, A1132 \\
    & Scalability and Large-Scale Processing & Highlights the ability of LLMs to support the processing and analysis of larger datasets or higher volumes of research material than would be feasible manually & A180, A767, A779, A877, A925, A966\\
    \SetCell[r=2]{c} Human-Level Alignment and Evaluation Quality & Evaluation Reliability and Measurement Quality & Captures the contribution of LLMs to improving the consistency, robustness, and overall quality of evaluation procedures and measurement activities & A63, A140, A161, A445, A776, A779, A925, A926, A1060 \\
    & Human-Level Agreement and Decision Concordance & Reflects the benefit of LLM outputs approximating, matching, or meaningfully aligning with human decisions, annotations, or expert judgments & A46, A105, A440, A767, A779, A926, A936, A1132 \\
    \SetCell[r=2]{c} Methodological Enhancement and Research Augmentation & Human-AI Collaboration and Workflow Augmentation & Represents the value of LLMs as complementary assistants that enhance researcher performance, support decision-making, and improve the execution of research workflows & A136, A227, A751 \\
    & Experimental Validity and Bias Control & Captures the potential of LLMs to strengthen experimental procedures, reduce sources of bias, and improve the methodological rigor of the study & A161, A1004 \\
\end{longtblr}

Among these, \textit{analytical performance and capability} emerges as the most prominent theme (23 instances), underscoring that the primary value attributed to LLMs lies in their ability to perform analytical tasks, such as classification, information extraction, and semantic reasoning. This theme is strongly supported by lower-level categories such as \textit{semantic understanding and reasoning capability} (10 instances) and \textit{automated classification performance} (8 instances), indicating that researchers frequently rely on LLMs for tasks that require abstraction, interpretation, and categorization of unstructured data. Additional contributions, such as \textit{information extraction and content structuring} (2 instances) and \textit{content filtering and relevance detection}, although less frequent (3 instances), further reinforce the perception of LLMs as analytical instruments.

The themes of \textit{human-level alignment and evaluation quality} (19 instances) and \textit{efficiency gains and scalability in empirical workflows} (16 instances) appeared with comparable frequency, suggesting that LLM adoption is motivated by both productivity gains and perceived output quality. Regarding efficiency, the lower-level themes of \textit{efficiency gains and workload reduction} (9 instances) and \textit{scalability and large-scale processing} (7 instances) indicate that LLMs are valued for accelerating research processes and handling large volumes of data with limited human intervention. Regarding quality, \textit{evaluation reliability and measurement quality} (10 instances) and \textit{human-level agreement and decision concordance} (9 instances) suggest that researchers perceive LLM outputs as sufficiently aligned with human judgment to support evaluative tasks. This balance between efficiency and perceived quality highlights a key motivation for LLM adoption: the promise of achieving human-like analytical outcomes at machine scale.

Finally, \textit{methodological enhancement and research augmentation} (6 instances) captures a more integrative role of LLMs within the research process. Lower-level themes such as \textit{human-AI collaboration and workflow augmentation} (4 instances) and \textit{experimental validity and bias control} (2 instances) suggest that LLMs are not only used as standalone tools but are increasingly embedded within research workflows to support study design, assist decision-making, and potentially mitigate certain methodological biases. However, the comparatively lower frequency of this theme indicates that such integrative uses are still emerging and less explored than efficiency-driven applications.

\begin{figure}[ht]
    \centering
    \includegraphics[width=\linewidth]{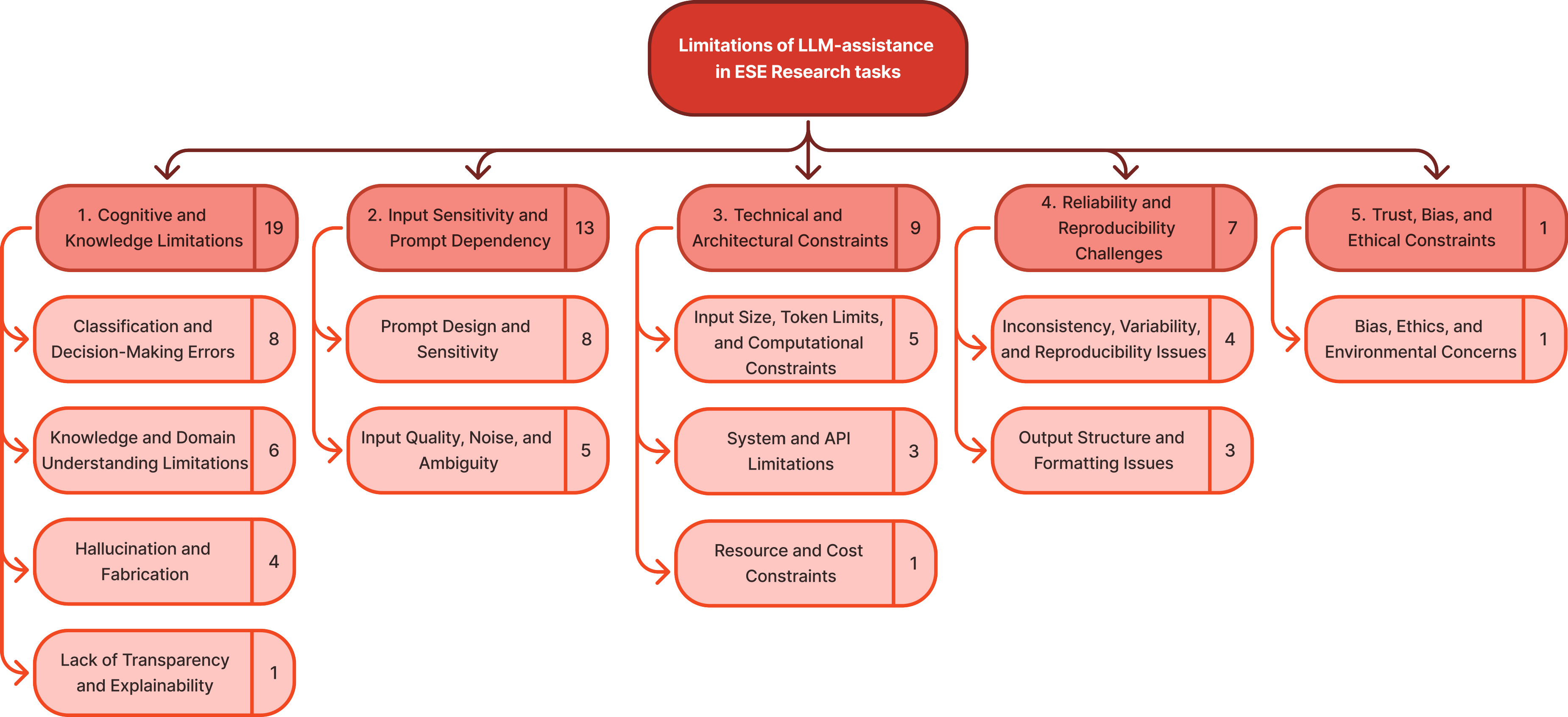}
    \caption{Thematic map of LLM limitations in software engineering research.}
    \label{fig:llm-limitations}
\end{figure}

\textbf{Turning to limitations}, the identified challenges can be grouped into five higher-level themes, as illustrated in Figure \ref{fig:llm-limitations}. These themes were supported by a total of 49 coded instances extracted from 21 out of the 50 selected papers; the remaining 29 papers did not report any limitations. The five higher-level themes are defined as follows:
\begin{enumerate}
    \item \textit{Cognitive and Knowledge Limitations} captures shortcomings in the ability of LLMs to correctly understand, reason about, and make decisions on research data, including errors related to classification, domain understanding, hallucination, and lack of explainability;
    \item \textit{Input Sensitivity and Prompt Dependency} reflects the extent to which LLM performance depends on how inputs are formulated, including sensitivity to prompt design, ambiguity in the provided material, and variability caused by noisy or incomplete inputs;
    \item \textit{Technical and Architectural Constraints} highlights limitations imposed by the underlying infrastructure of LLM use, such as token and input-size restrictions, system or API limitations, and computational or cost-related barriers;
    \item \textit{Reliability and Reproducibility Challenges} emphasizes difficulties in obtaining stable, consistent, and reproducible outputs, as well as problems related to output structure, formatting, and variability across executions;
    \item \textit{Trust, Bias, and Ethical Constraints} reflects concerns about whether LLM outputs can be used responsibly and fairly, particularly in light of risks related to bias, ethical issues, and broader societal or environmental implications.
\end{enumerate}

The lower-level themes associated with each higher-level category, together with their definitions, are presented in Table \ref{tab:llm_limitation_themes}.

The most dominant theme is \textit{cognitive and knowledge limitations} (19 instances), indicating that concerns about the fundamental capabilities of LLMs remain prevalent. Within this theme, \textit{classification and decision-making errors} (8 instances), \textit{knowledge and domain understanding limitations} (6 instances), and \textit{hallucination and fabrication} (4 instances) are particularly salient, pointing to persistent issues in accuracy, factual reliability, and domain specificity. Additionally, \textit{lack of transparency and explainability} (1 instance) highlights the challenges researchers face in interpreting and trusting model outputs, especially in contexts requiring accountability and rigor.

The second most prominent theme, \textit{input sensitivity and prompt dependency} (13 instances), reflects the extent to which LLM performance is contingent on how tasks are formulated. The relatively high frequency of \textit{prompt design and sensitivity} (8 instances), compared to \textit{input quality, noise, and ambiguity} (5 instances), suggests that prompt engineering itself has become a critical and non-trivial aspect of using LLMs effectively. This reliance introduces an additional layer of complexity, as small variations in input formulation can lead to substantially different outputs, thereby undermining consistency.

\begin{longtblr}[
    caption={Definitions of the lower-level themes and their relationship to the higher-level limitation themes.},
    label={tab:llm_limitation_themes}
]{
    colspec = {Q[c, m, 2.45cm] Q[j, m, 2.95cm] X[j, m] Q[l, m, 1.8cm]},
    hlines,
    hline{14} = {1pt, solid},
    row{1} = {bg=black, fg=white, font=\bfseries},
    rows = {font=\footnotesize},
    rowsep = 1pt
}
    \textbf{Higher-level theme} & \textbf{Lower-level themes} & \textbf{Description} & \textbf{Paper Ids}  \\
    \SetCell[r=4]{c} Cognitive and Knowledge Limitations & Classification and Decision-Making Errors & Captures limitations arising when LLMs misclassify instances, make incorrect judgments, or produce decisions that are not sufficiently reliable for the empirical task at hand & A136, A171, A447, A925, A936, A1870 \\
    & Knowledge and Domain Understanding Limitations & Reflects shortcomings in the model's ability to adequately understand domain-specific concepts, contextual nuances, or specialized software engineering knowledge required for the research task & A136, A751, A776, A1132 \\
    & Hallucination and Fabrication & Represents the risk of LLMs generating unsupported, fabricated, or factually incorrect content, which can compromise the validity of empirical results & A227, A751, A925, A1673 \\
    & Lack of Transparency and Explainability & Highlights the difficulty of understanding why the model produced a given output, limiting researchers' ability to interpret, justify, or critically assess LLM-assisted decisions & A136  \\
    \SetCell[r=2]{c} Input Sensitivity and Prompt Dependency & Prompt Design and Sensitivity & Captures the dependence of LLM performance on how prompts are formulated, including the risk that small variations in wording, structure, or level of detail can lead to different outcomes & A64, A96, A105, A134, A925, A926, A1132  \\
    & Input Quality, Noise, and Ambiguity & Reflects limitations caused by noisy, incomplete, ambiguous, or low-quality input material, which can reduce the accuracy and usefulness of LLM outputs & A140, A925, A926, A1132 \\
    \SetCell[r=3]{c} Technical and Architectural Constraints & Input Size, Token Limits, and Computational Constraints & Represents barriers imposed by context-window limits, token restrictions, hardware requirements, or computational constraints that hinder the application of LLMs in empirical workflows & A64, A136, A447, A1354, A1870 \\
    & System and API Limitations & Captures problems stemming from the surrounding technical infrastructure, such as unavailable features, API restrictions, service instability, or integration challenges & A105, A140 \\
    & Resource and Cost Constraints & Reflects the financial and resource-related burden of using LLMs, including API costs, infrastructure expenses, and the effort required to support large-scale or repeated use & A140 \\
    \SetCell[r=2]{c} Reliability and Reproducibility Challenges & Inconsistency, Variability, and Reproducibility Issues & Captures the difficulty of obtaining stable and reproducible outputs, including variation across runs, sensitivity to configuration, and inconsistency in model behavior & A134, A1109, A1354 \\
    & Output Structure and Formatting Issues & Reflects limitations related to outputs that are poorly structured, difficult to parse, incorrectly formatted, or otherwise unsuitable for direct use in empirical analysis pipelines & A779, A1004 \\
    \SetCell[r=1]{c} Trust, Bias, and Ethical Constraints & Bias, Ethics, and Environmental Concerns & Represents broader concerns about the responsible use of LLMs, including biased outputs, ethical risks, and potential environmental costs associated with model development and use & A925 \\
\end{longtblr}

\textit{Technical and architectural constraints} (9 instances) and \textit{reliability and reproducibility challenges} (7 instances) further emphasize practical limitations that affect the robustness of LLM-based research. Constraints such as \textit{input size and token limits} (5 instances), \textit{system and API limitations} (3 instances), and \textit{resource and cost constraints} (1 instance) indicate that the deployment of LLMs is not only a methodological concern but also a technical and infrastructural one. Meanwhile, issues related to \textit{inconsistency, variability, and reproducibility} (4 instances), along with \textit{output structure and formatting issues} (3 instances), reveal that even when LLMs perform adequately, their outputs may lack the stability and standardization required for rigorous empirical research.

In contrast, \textit{trust, bias, and ethical constraints} are comparatively underrepresented (1 instance), despite their recognized importance. This limited reporting may suggest that ethical considerations, including \textit{bias, fairness, and environmental impact}, are either underexplored or insufficiently documented in the current body of literature, rather than being negligible concerns.

As previously mentioned, a notable proportion of studies did not explicitly report benefits (17 out of 50) or limitations (29 out of 50), suggesting that LLMs are often incorporated without systematic reflection on their implications. This omission points to a tendency to treat LLMs as auxiliary instruments embedded within the research pipeline, rather than as methodological components whose impact warrants explicit and critical evaluation.

Moreover, the benefits reported in the literature are often not derived from observations within the studies themselves, but are instead introduced as a priori justifications for adopting LLMs. In many cases, advantages such as efficiency, scalability, or human-level agreement are presented as motivating assumptions, drawing implicitly or explicitly on prior work, rather than as outcomes empirically demonstrated in the specific research context. As a result, the role of LLMs in these studies is frequently justified based on expected capabilities, rather than on evidence generated through their actual application within the study.

An additional insight emerges when considering the interplay between reported benefits and limitations across the literature. Several benefits identified in some studies are directly challenged by limitations reported in others, revealing underlying tensions in how LLMs are perceived and utilized. For instance, the frequently cited benefit of \textit{human-level alignment and evaluation quality} contrasts with reported issues of \textit{inconsistency, variability, and reproducibility}, suggesting that while LLM outputs may approximate human judgments in isolated cases, they may fail to do so reliably across repeated runs or slightly varied inputs. 

Similarly, claims regarding efficiency gains and scalability are counterbalanced by technical and architectural constraints, including token limits, computational costs, and API dependencies, which can restrict scalability in practice. Likewise, the perceived strength of analytical performance and reasoning capability is tempered by well-documented issues such as hallucinations, classification errors, and limited domain understanding, indicating that apparent task performance does not necessarily equate to robust or trustworthy outcomes.

These results reinforce the view that the field remains in an exploratory and formative stage. While researchers actively leverage LLM capabilities, there is a need to more systematically align claimed benefits with context-specific empirical validation. Advancing the maturity of this research area will require more rigorous and context-aware experimental designs, clearer reporting of assumptions and transferability conditions, and evaluation protocols that explicitly examine both the strengths and the limitations of LLMs within the specific settings in which they are applied.

\begin{tcolorbox}[title=Summary of RQ4, fonttitle=\small\bfseries, colframe=black, fontupper=\small]
\begin{itemize}[leftmargin=*, itemsep=2pt, topsep=2pt]
\item Reported benefits emphasize efficiency, scalability, and analytical capability.
\item Benefits are often assumed rather than empirically validated.
\item Limitations include hallucinations, inconsistency, and prompt sensitivity.
\item Reproducibility challenges emerge as a central concern.
\end{itemize}
\end{tcolorbox}

\subsection{RQ5 - To what extent do studies proposing LLM-assisted empirical software engineering research report the information required to enable reproducible use of LLMs?}

To assess whether LLM-assisted empirical studies provide sufficient information to support reproducibility, we analyzed nine transparency-related reporting elements derived from existing methodological guidelines~\cite{wagner2025guidelines}. These elements are summarized in Table~\ref{tab:assessment-questions}.

\begin{figure}[ht]
    \centering
    \includegraphics[width=\linewidth]{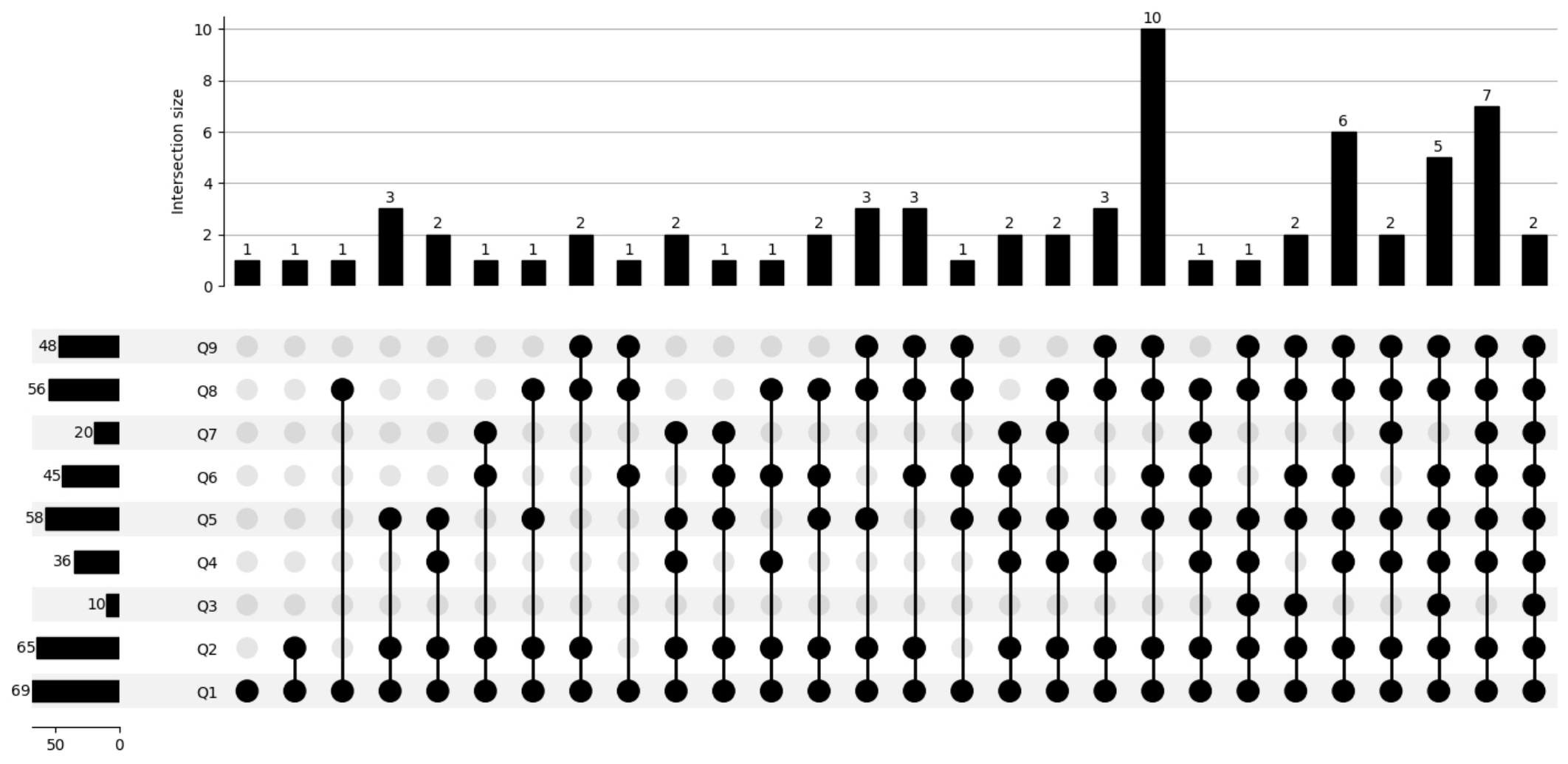}
    \caption{UpSet plot showing the frequency and co-occurrence of transparency and reproducibility reporting practices (Q1–Q9) across identified LLM-assisted ESE research tasks.}
    \label{fig:upsetplot-assessment}
\end{figure}

Figure~\ref{fig:upsetplot-assessment} presents an UpSet plot illustrating the frequency and co-occurrence of transparency and reproducibility reporting practices across LLM-assisted ESE tasks. In total, 69 LLM-assisted tasks were identified across the analyzed papers. Because a single study may report multiple LLM-assisted tasks, the unit of analysis in this figure is the task rather than the study, which explains why the number of observations exceeds the number of papers. 

The UpSet plot summarizes both the frequency of individual reporting practices and the combinations in which they appear across tasks. The horizontal bars indicate how many tasks satisfy each reporting question (Q1-Q9), while the vertical bars represent intersections, i.e., the number of tasks simultaneously satisfying the reporting practices indicated by the connected dots below.

Across the analyzed studies, the results indicate that basic identification details of the LLMs used, such as the specific model version, are consistently reported. In contrast, more detailed methodological information, such as prompts, experimental settings, and validation procedures, is documented less consistently.

The most frequently reported element is Q1 (declaration of LLM use and its role), which appears in all tasks. This is expected, since the selection criteria for this study intentionally included only papers that explicitly employ LLMs to support at least one research task. Information about the specific model version (Q2) is also widely reported, appearing in 65 tasks. Reporting of prompts (Q5) and the use of human validation for LLM outputs (Q8) are also relatively common. Regarding the reporting of model configuration (Q4), although more than half of the analyzed tasks provide this information, 11 report it exclusively in the replication package, without discussing in the paper the rationale behind the configuration decisions.

In contrast, several reporting practices remain uncommon. For example, Q3 (reporting the experiment date) and Q7 (using an open LLM as a baseline) appear considerably less often, indicating limited disclosure of temporal context and baseline comparisons. Similarly, although some studies report performing human validation (Q8), fewer explicitly describe how the validation process was conducted (Q9), suggesting that the validation methodology itself is often underreported.

The intersection analysis further highlights how transparency practices co-occur across tasks. The largest intersections involve combinations of Q1, Q2, Q5, Q6, and Q8, representing tasks that provide relatively comprehensive documentation of LLM usage and prompt-related information. Nevertheless, the presence of many smaller intersections indicates substantial variability across studies, reflecting heterogeneous and often incomplete documentation of LLM-assisted experimental settings.

In summary, these results highlight two main reporting patterns. First, several methodological elements required for reproducibility, such as prompts, configuration parameters, and validation procedures, are reported inconsistently across studies. Second, comprehensive documentation covering all transparency elements remains rare. As a result, many methodological details necessary to reproduce LLM-assisted empirical analyses are either partially documented or omitted altogether.

Importantly, the studies analyzed in this review originate from venues that traditionally emphasize empirical rigor and methodological transparency. From this perspective, the observed inconsistencies in reporting practices are particularly noteworthy. If researchers working within the ESE community, who are expected to be familiar with established practices for experimental reporting and reproducibility, still omit essential details about LLM usage, it is likely that reporting practices in the broader body of software engineering research applying LLMs are even less systematic. Consequently, the transparency gaps identified in this study may reflect a broader methodological challenge in the emerging literature on LLM-supported software engineering research.

These findings also have implications for the peer-review process. The presence of reporting gaps in papers published in established software engineering venues suggests that such omissions were not consistently identified during peer review. This likely reflects the novelty of LLM-assisted empirical research rather than shortcomings of individual reviewers. Because the methodological implications of integrating LLMs into empirical workflows are still being understood, the community is still developing shared expectations regarding what information must be reported to ensure transparency and reproducibility. As a result, both authors and reviewers may currently lack consolidated guidelines for evaluating the methodological completeness of LLM-assisted empirical studies.

These findings are strongly aligned with recent calls for methodological guidelines for LLM-assisted empirical research~\cite{baltesllm_guidelines}. In particular, the observed limitations related to hallucinations, prompt sensitivity, and incomplete reporting reinforce the need for standardized practices to ensure transparency, reproducibility, and reliability. Our results provide empirical evidence supporting these concerns, highlighting that many of the risks identified in prior guideline-oriented work are already manifest in current research practices.

\begin{tcolorbox}[title=Summary of RQ5, fonttitle=\small\bfseries, colframe=black, fontupper=\small]
\begin{itemize} [leftmargin=*, itemsep=2pt, topsep=2pt]
\item Reporting practices for LLM-assisted studies are often incomplete.
\item Key elements such as prompts, configurations, and validation procedures are frequently missing.
\item This lack of transparency limits reproducibility and comparability.
\item The findings underscore the need for standardized reporting practices.
\end{itemize}
\end{tcolorbox}

\section{Discussion}
\label{sec:discussion}
This section synthesizes the findings across RQ1-RQ5 to provide an integrated interpretation of how LLMs are currently employed as methodological instruments in empirical software engineering, and the implications this has for research practice. Rather than discussing each research question in isolation, we examine the interconnections between observed usage patterns, their methodological consequences, and prevailing reporting practices. 

\subsection{Automation-centered Integration of LLMs}
A consistent pattern emerges when jointly analyzing RQ1 and RQ2: the phases of empirical research supported by LLMs are closely aligned with the roles they predominantly assume. The concentration of LLM usage in MSR and controlled experiments, particularly in data processing and analysis phases, directly corresponds to the prevalence of roles such as \textit{semantic classifier} and \textit{criteria-based screener}. These roles encapsulate tasks that can be expressed as structured input-output mappings, where decision boundaries are relatively well-defined.

This alignment is further reinforced by the dominance of \textit{automation} as the primary mode of LLM involvement, as seen in RQ3. Rather than supporting human reasoning processes, LLMs are frequently deployed as end-to-end mechanisms that execute clearly specified tasks. This suggests that current adoption is driven by the ability to operationalize LLMs within controlled settings, where their probabilistic behavior can be constrained through task design. Consequently, LLM integration in ESE workflows appears to be shaped less by their full theoretical capabilities and more by the feasibility of embedding them into structured, low-ambiguity tasks.

This pattern can be interpreted as a form of constrained adoption, in which LLMs are preferentially integrated into contexts where their non-deterministic behavior can be bounded and evaluated. It also aligns with the benefits identified in RQ4, particularly efficiency gains and analytical performance, which are most readily realized in tasks with clearly defined outputs and evaluation criteria.

\subsection{From Decision Production to Limited Decision Support}
The roles identified in RQ1 further clarify how LLMs are operationalized within these automated workflows. The most frequent roles, such as \textit{semantic classifier} and \textit{criteria-based screener}, involve assigning labels or making rule-based decisions over inputs. These roles indicate that LLMs are frequently used to directly produce decisions in structured tasks.

In contrast, the \textit{decision support} category is minimally represented, indicating that LLMs are rarely used to assist researchers in making decisions through explanations, alternatives, or interactive feedback. Instead, their outputs are typically consumed as final or intermediate results within the workflow.

This reveals a fundamental characteristic of current LLM integration: rather than reshaping how decisions are made in empirical research, LLMs are primarily used to replace decision-making steps altogether. In other words, current usage is oriented toward \textit{decision production} rather than \textit{decision support}. Notably, this pattern contrasts with the benefits identified in RQ4 related to human-AI collaboration and workflow augmentation. While such benefits suggest a collaborative paradigm, the observed usage patterns indicate that LLMs are rarely embedded in interactive decision processes. This reveals a gap between the conceptual framing of LLMs as collaborative tools and their practical use as autonomous task executors.

\subsection{Concentration on Structured Tasks and Capability-Risk Alignment}
The concentration of roles around classification, filtering, and evaluation tasks indicates that LLM adoption is currently focused on activities that are structured and operationalizable through clear input-output mappings. These tasks are naturally amenable to automation, as they involve bounded outputs such as labels, scores, or rankings.

This pattern is consistent with the distribution of tasks across research methods observed in RQ2. Methods that rely on large-scale, structured data, such as mining software repositories, show higher levels of LLM support, whereas methods that depend more heavily on qualitative reasoning or interpretive judgment exhibit limited adoption.

Beyond reflecting capability, this distribution can be interpreted as a form of capability-risk alignment. LLMs are preferentially applied to tasks where errors are easier to detect, quantify, or mitigate, and where their impact on the overall validity of the study is limited. Conversely, tasks requiring deeper contextual interpretation or domain-specific reasoning are less explored, as errors in these settings are harder to identify and may compromise research outcomes.

This selective adoption is consistent with the limitations identified in RQ4, particularly hallucinations, domain knowledge constraints, and lack of explainability. As a result, current LLM usage reflects not only what these models can do, but also where their limitations are most manageable from a methodological standpoint.

\subsection{Uneven Adoption across the Research Lifecycle}
The mapping of tasks to standardized research phases reveals a clear asymmetry in how LLMs are integrated across the empirical research lifecycle. As shown in RQ2, LLM-supported tasks are concentrated in \textit{data processing} and \textit{analysis \& synthesis}, while their presence in \textit{planning \& design} and \textit{reporting} phases remains limited.

This distribution indicates that LLMs are primarily integrated into operational phases where tasks can be formalized and scaled, rather than into epistemic phases that involve study design, conceptual reasoning, or interpretation of results. Such phases require justification, contextual awareness, and theoretical grounding, which are less compatible with the current strengths and limitations of LLMs.

Additionally, LLM integration in later stages of the research lifecycle (\textit{reporting}) is either limited or not yet systematically documented.

\subsection{Transparency and Reproducibility in LLM-assisted Workflows}
The results of RQ5 introduce an important methodological dimension to these observations. Our analysis shows that many studies do not provide sufficient detail to enable the reproducibility of LLM-assisted tasks. Key elements such as model versions, prompts, configuration parameters, and details of human validation are often incompletely reported or omitted.

When considered alongside the findings from RQ3, this lack of transparency raises important concerns. As LLMs are frequently used in automated settings, their outputs become integral components to the research process. In such cases, insufficient reporting of how these outputs are generated limits the ability to reproduce, interpret, and critically assess the results.

More fundamentally, LLMs introduce a hidden methodological layer into empirical workflows. Decisions embedded in prompts, parameter settings, and interaction strategies directly influence outcomes, yet are often not treated as first-class methodological elements. This obscures important aspects of the research process and hinders both reproducibility and accountability.

 This issue is further exacerbated by the predominance of automation. When LLM outputs are directly incorporated into research pipelines without sufficient transparency, the lack of reproducibility becomes not only a reporting issue but a potential threat to the validity of empirical findings.

\subsection{Ecosystem Concentration and Its Implications}
The analysis of model usage reveals a strong concentration around a small set of LLMs, particularly ChatGPT. This indicates that much of the empirical evidence on LLM-assisted ESE is derived from a limited set of model families.

While this concentration may facilitate comparability across studies, it also implies that current research findings are shaped by the specific characteristics, limitations, and design choices of a narrow subset of models. As a result, observed practices may reflect not only methodological choices but also the design and accessibility of the most commonly used tools.

This raises important concerns regarding external validity. Findings derived from a limited set of models may not generalize to alternative architectures, configurations, or future generations of LLMs. Moreover, when combined with inconsistent reporting practices, this concentration further limits the ability to disentangle whether observed outcomes are due to methodological design or model-specific behavior.

Our findings also provide an empirical foundation for recent guideline-oriented efforts, such as the work by Baltes et al.~\cite{baltesllm_guidelines}. While these efforts propose recommended practices for conducting LLM-assisted empirical studies, our results reveal that current usage only partially aligns with such recommendations. For example, we observe substantial gaps in reporting practices, particularly regarding prompts, configurations, and validation procedures, which are key elements emphasized in existing guidelines. This gap suggests that, although the community recognizes the need for methodological rigor, the adoption of standardized practices remains limited.

From this perspective, our study complements prior work by grounding guideline development in empirical evidence. Rather than proposing new guidelines, we identify where current practices fall short, thereby helping to inform the refinement, prioritization, and operationalization of future methodological recommendations.

\section{Research Agenda and Future Directions}
\label{sec:researchAgenda}
Building on the findings synthesized in the previous section, we outline a research agenda to guide the next phase of LLM integration into empirical software engineering. Current usage is largely centered on the automation of structured tasks, suggesting that LLM-assisted ESE is still in an early stage of methodological development. Advancing this area requires moving beyond task execution toward more collaborative, transparent, and methodologically robust forms of integration.

A key direction for future work is \textbf{to shift from automation-centered usage toward human-AI collaboration}. Our findings show that LLMs are primarily used to perform tasks with minimal human intervention, with limited support for interactive or iterative workflows. Future research should explore how LLMs can act as collaborative components, providing intermediate outputs, suggestions, or alternative perspectives that remain subject to human interpretation and validation. Such approaches can enable more dynamic forms of empirical analysis, where researchers and LLMs jointly contribute to the development of insights.

Closely related to this is the \textbf{need to enable decision support within empirical workflows}. While LLMs are frequently used to produce decisions in structured tasks, their use in assisting human decision-making remains limited. Future work should investigate how LLMs can support researchers in tasks involving ambiguity, interpretation, or synthesis of evidence. This includes developing mechanisms for generating explanations, communicating uncertainty, and presenting multiple candidate outputs, allowing researchers to make informed decisions rather than relying solely on automated results.

Another important direction is \textbf{to expand the range of tasks supported by LLMs}. Current applications are concentrated in classification, filtering, and evaluation, reflecting a narrow operationalization of LLM capabilities. Future research should explore how LLMs can support more cognitively demanding activities, such as reasoning, interpretation, and synthesis. This includes their potential use in supporting the formulation of research questions, assisting qualitative analysis, or contributing to the interpretation of empirical findings.

\begin{table*}
\caption{Summary of identified gaps and research directions for LLM-assisted empirical software engineering.}
\label{tab:research-agenda-summary}
\begin{tblr}{
  colspec = {Q[l,m,3cm] Q[l,m,2.2cm] Q[l,m,5.2cm] X[l,m]},
  hlines,
  row{1} = {bg=black, fg=white, font=\bfseries, ht=1.1cm},
  hline{11} = {1.5pt, solid},
  rows = {font=\scriptsize},
  rowsep = 1pt
}
\textbf{Gap Category} & \textbf{Evidence} & \textbf{Identified Gap} & \textbf{Research Directions} \\
Task Diversity & RQ1 & Narrow focus on classification, filtering, and evaluation tasks & Explore use of LLMs in reasoning, synthesis, and interpretation tasks \\
Agentic AI Integration & RQ1, RQ2, RQ3 & Current studies rely predominantly on direct LLM usage, with little exploration of agentic systems capable of planning, orchestration, and tool use & Investigate agentic AI support for empirical workflows by evaluating planning capabilities, tool integration, and multi-step execution \\
Discovery Support & RQ1, RQ2, RQ3 & Current studies focus mainly on structured task execution, with little exploration of LLMs or agents as support for hypothesis generation, exploratory reasoning, and discovery-oriented inquiry & Investigate how LLMs and agents can assist hypothesis formation, exploratory analysis, alternative explanation generation, and discovery-oriented ESE workflows\\
Lifecycle Coverage & RQ2 & Uneven adoption across research phases; limited support for planning, data collection, and reporting & Investigate LLM support for study design, instrument generation, and reporting assistance \\
Human--AI Collaboration & RQ3 & LLMs are primarily used for automation, with limited support for collaborative or interactive workflows & Design human-in-the-loop systems; develop interactive and iterative LLM workflows \\
Decision Support & RQ3 & Minimal use of LLMs to support human decision-making; focus is on decision production & Develop explainable LLM outputs; support uncertainty communication and alternative suggestions \\
Evaluation of Human--LLM Interaction & RQ3, RQ4 & Limited evaluation of how LLMs affect researcher behavior and decision quality & Develop metrics for trust, cognitive load, and interaction quality \\
Transparency and Reproducibility & RQ5 & Incomplete reporting of prompts, configurations, and model details & Establish reporting standards; define reproducibility guidelines for LLM-assisted studies \\
Model Ecosystem Dependence & Section 4 & Strong reliance on a small set of proprietary LLMs (e.g., ChatGPT) & Evaluate diverse models; include open-source alternatives; benchmark across LLMs \\
\end{tblr}
\end{table*}

An especially relevant direction is to investigate the role of \textbf{agentic AI in empirical software engineering}. Most of the studies identified in this review rely on standalone LLM usage, in which the model is prompted to perform a specific task in isolation. By contrast, agentic systems can support more complex assistance by planning multi-step actions, invoking LLMs iteratively, and interacting with external tools or data sources. This makes them particularly relevant for empirical workflows that require task decomposition, intermediate decision-making, and coordination across multiple research activities. Future research should therefore examine how agentic AI can extend current forms of LLM support in ESE, while also characterizing its limitations, including issues of reliability, transparency, controllability, cost, and error propagation across multi-step workflows.

A related and still underexplored direction is the potential of \textbf{LLMs and agentic AI to support hypothesis formation and discovery in empirical software engineering}. Recent work on LLMs in the scientific method suggests that these systems may contribute not only to task execution, but also to the generation of novel hypotheses, the exploration of broader hypothesis spaces, and the coordination of iterative hypothesis $\rightarrow$ experiment $\rightarrow$ observation loops, especially when combined with planning and tool-use capabilities \cite{Zhang2025}. A similar opportunity exists in ESE, where researchers also seek to uncover valid explanations, regularities, and causal relationships, not about natural phenomena, but about software engineering practices, processes, artifacts, and human activities. Future work should therefore investigate how LLMs and agents can assist ESE researchers in formulating hypotheses, identifying promising research directions, suggesting alternative explanations, and supporting exploratory analyses aimed at surfacing previously overlooked patterns.

In addition, our results indicate that \textbf{LLM adoption is uneven across the empirical research lifecycle}, with limited support for planning, data collection, and reporting phases. Expanding LLM integration into these underexplored phases represents a promising direction. For instance, LLMs may assist in refining research designs, generating or validating data collection instruments, or supporting the structured reporting of empirical studies. In the reporting phase, particular attention should be given to ensuring transparency regarding how LLMs contribute to the generation of research outputs.

\textbf{Improving transparency and reporting practices} is also critical for the advancement of LLM-assisted ESE. Our findings show that key details required for reproducibility, such as model versions, prompts, and configuration parameters, are often incompletely reported. While recent work \cite{baltesllm_guidelines} has already proposed guidelines for reporting LLM usage, our results suggest that these practices have not yet been consistently adopted or operationalized within the community. Future efforts should therefore focus not only on refining existing guidelines, but also on reinforcing their use and promoting their widespread adoption. This includes clarifying expectations for documenting LLM-assisted workflows, integrating reporting standards into publication and review processes, and encouraging authors to treat LLM configurations and prompts as first-class methodological elements. Strengthening and disseminating these practices can improve reproducibility, facilitate comparison across studies, and support more rigorous and transparent evaluation of LLM-based methods.

Furthermore, the strong concentration of studies around a small set of LLMs suggests the \textbf{need to diversify the models used in empirical research}. Future studies should investigate a broader range of LLM architectures to better understand how model characteristics influence empirical outcomes. This can help reduce dependence on specific tools and improve the generalizability of findings.

Finally, as LLMs become more integrated into empirical workflows, there is a \textbf{need to evaluate not only task performance but also the nature of human-LLM interaction}. Future research should examine how LLMs affect the research process, including aspects such as researcher effort, decision quality, and interaction patterns. Understanding how researchers interpret, validate, and trust LLM outputs is essential to ensure that these systems are used in a way that supports, rather than undermines, rigorous empirical research.

Table~\ref{tab:research-agenda-summary} summarizes the key gaps identified across RQ1-RQ5 and outlines corresponding research directions grounded in our findings.

\section{Threats to Validity}
\label{sec:threats}

As with any systematic literature review, this study is subject to several threats to validity. To mitigate these threats, we followed a predefined and documented review protocol and adopted established guidelines for conducting secondary studies in empirical software engineering \cite{kitchenham2002preliminary,ExperimentationSEWohlin}. Nevertheless, some limitations remain and are discussed below.

\textbf{Construct validity} concerns whether the study accurately captures the phenomenon under investigation, namely the use of LLMs to support empirical software engineering activities. A primary threat arises from the operationalization of \emph{LLM-assisted ESE}. Although the protocol defines explicit inclusion and exclusion criteria, relevant studies may describe LLM usage implicitly, employ alternative terminology, or treat LLM support as a secondary methodological component. Such cases may be difficult to identify through title and abstract screening alone, since LLM usage to support empirical research tasks is often described only in methodological sections. To mitigate this threat, we did not rely solely on titles, abstracts, or database metadata. Instead, we applied a keyword-based filtering procedure to the full text of each paper in the selected proceedings and journal issues. Papers containing at least one predefined LLM-related keyword were then manually inspected to determine whether the LLM was actually used to support or automate an empirical software engineering research task. This combination of full-text keyword filtering and manual inspection helped reduce the risk of missing relevant studies whose LLM usage was not explicit in the title, abstract, or author-provided keywords.

Another construct validity threat relates to the classification of ESE research tasks, LLM roles, and reported benefits or limitations. These constructs often require interpretive judgment, particularly when primary studies provide limited methodological detail. To reduce ambiguity, the protocol specifies a structured data extraction form aligned with the research questions and the use of controlled vocabularies to normalize categorical fields.

\textbf{Internal validity} refers to potential bias introduced during study selection, data extraction, and synthesis. Researcher bias may affect inclusion decisions or coding judgments, especially in borderline cases or when interpreting qualitative descriptions of LLM usage. To mitigate this threat, two researchers first independently screened and extracted data from a subset of studies. This calibration phase was used to align their interpretation of the inclusion and exclusion criteria, refine the extraction form, and establish consistent coding procedures. Disagreements identified during this phase were discussed and resolved by consensus. After calibration, the remaining studies were divided between the two researchers, and each researcher independently screened and extracted data from their assigned subset. Ambiguous or uncertain cases identified during the full selection and extraction process were jointly discussed and resolved through consensus, with third-party arbitration available if necessary.

Another internal validity threat stems from incomplete or ambiguous reporting in primary studies. Information regarding prompt design, model configuration, or the degree of human involvement is often underspecified. Rather than inferring missing details, we adopted conservative coding decisions and recorded only information explicitly reported in the studies, using notes to document uncertainty where appropriate.

\textbf{External validity} concerns the generalizability of the findings beyond the reviewed studies. The scope of this review is restricted to peer-reviewed publications from a curated set of established software engineering conferences and journals between 2020 and 2025. Venue selection was guided by objective quality and scope criteria, including recognized CORE A*/A classifications and explicit emphasis on empirical software engineering research. This restriction was intentional, as the goal of the review is to synthesize evidence on the methodological use of LLMs within empirical software engineering workflows, contributions that are most likely to appear in rigorous archival venues.

While this strategy enhances methodological consistency, reproducibility, and quality control, relevant work published in workshops, emerging venues, preprints, or outside the selected time window may not have been captured. The threat is partially mitigated by including a broad range of high-impact venues spanning empirical software engineering, mining software repositories, and AI engineering. Nonetheless, the findings should be interpreted as reflecting the state of LLM-assisted empirical software engineering research within established scholarly outlets during the selected period, rather than as an exhaustive or future-proof account of all possible applications.

\textbf{Reliability} refers to the consistency and repeatability of the review process. Subjectivity in qualitative coding and narrative synthesis represents a potential threat, as different researchers may categorize studies differently. The protocol mitigates this threat through independent data extraction by multiple researchers, structured extraction forms, and calibration through pilot testing. Detailed documentation of constructs, decision rules, and extracted data supports transparency and facilitates replication or extension of the review.

\textbf{Conclusion validity} concerns the soundness of the inferences drawn from the synthesized evidence. A key threat arises from the heterogeneity of the primary studies, which vary in empirical methods, datasets, LLM configurations, and evaluation strategies. Drawing causal claims across such diverse studies would be inappropriate. To mitigate this threat, the analysis emphasizes descriptive and classificatory synthesis rather than causal inference. Reported benefits, limitations, and threats to validity are treated as claims made in the primary studies, not as independently validated outcomes. Frequency analysis and narrative synthesis are used to identify patterns while avoiding overinterpretation.

Publication bias also represents a potential threat, as studies reporting positive outcomes may be more likely to be published. This limitation is explicitly acknowledged when interpreting the results.

Finally, this review faces \textbf{threats specific to studying LLM-assisted research}. Primary studies often lack detailed reporting on prompts, model versions, parameter settings, and the degree of human oversight, which limits reproducibility and comparability. These inconsistencies complicate cross-study synthesis and the assessment of methodological rigor. Rather than attempting to normalize or infer missing information, this review systematically identifies and documents validity threats associated with the use of LLMs as reported by the primary studies. As a result, transparency gaps and reporting limitations are treated as empirical observations and discussed as part of the study’s contributions.

\section{Related Work}
\label{sec:relatedwork}

This section positions the present review with respect to prior secondary studies in software engineering and related areas. We focus on (1) existing surveys and systematic literature reviews on large language models (LLMs) in software engineering, which primarily study LLMs as objects of empirical evaluation, and (2) prior work on automation and machine-learning support for empirical software engineering. We then clarify how this review advances beyond existing work.

\subsection{Surveys and Reviews on LLMs in Software Engineering}

A growing body of surveys and systematic literature reviews has examined the use of LLMs in software engineering. These studies synthesize evidence across tasks such as code generation, program repair, testing, documentation, requirements analysis, and security~\cite{LLM4SE_LITREVIEW,zhou2022survey,Chen2021}, typically organizing the literature by task categories, model architectures, prompting strategies, datasets, and evaluation metrics, with a strong emphasis on benchmarking performance.

While this line of work has been instrumental in consolidating knowledge about the capabilities and limitations of LLMs for software engineering practice, it predominantly treats LLMs as objects of empirical evaluation rather than as instruments supporting the research process. Consequently, methodological aspects, such as how LLMs are integrated into empirical workflows, how human oversight is exercised, and how their use impacts validity and transparency, remain largely unexplored.

More recently, research has begun to investigate LLMs as evaluation mechanisms for software engineering artifacts. He et al.~\cite{LLMAsAJudge4SE} provide a comprehensive review of the \textit{LLM-as-a-Judge} paradigm, in which LLMs are used to assess the quality of artifacts such as generated code, requirements, and test cases. Their study highlights the potential of LLMs for scalable, multi-criteria evaluation, while also identifying key challenges, including inconsistency, prompt sensitivity, and the lack of robust evaluation frameworks.

However, this perspective remains centered on artifact evaluation. In contrast, our work adopts a complementary view by examining how LLMs function as methodological instruments within empirical software engineering research. Rather than focusing on what LLMs evaluate, we analyze how they are operationalized across research tasks and phases, providing a task-centric and cross-method characterization of LLM-assisted empirical workflows.

\subsection{Automation and Machine Learning Support for Empirical Software Engineering}

Automation has long been employed to scale empirical software engineering research. Foundational work on mining software repositories enabled large-scale empirical studies of version control systems, issue trackers, and developer communication artifacts \cite{Hassan2008,Zimmermann2005}. Subsequent research introduced natural language processing and machine-learning techniques to support tasks such as defect prediction, topic modeling, and traceability analysis \cite{Hindle2012}.

Methodological guidelines and secondary studies in this area highlight the benefits of automation for scalability and consistency, while also noting challenges related to task specificity, feature engineering, and interpretability \cite{Shull2008,ExperimentationSEWohlin}. Traditional automation approaches typically rely on narrowly scoped models designed for specific empirical tasks and require substantial domain expertise to configure and validate.

LLMs represent a qualitative departure from these earlier approaches. Unlike task-specific models, LLMs can operate across heterogeneous artifacts and multiple stages of the empirical research lifecycle, including activities that require semantic interpretation and abstraction. At the same time, their probabilistic nature and limited transparency introduce new methodological risks, such as hallucinations, bias, and threats to reproducibility, which are not well addressed by existing work on automated empirical analysis.

Despite growing interest in computational support for ESE, prior research has not yet provided a systematic synthesis of how LLMs are adopted as general-purpose research assistants.

\subsection{Advancement Beyond Prior Work}

The present systematic literature review advances the state of the art by shifting the analytical focus from LLMs as objects of empirical evaluation to LLMs as methodological instruments used in empirical software engineering research. To the best of our knowledge, this is the first review to systematically synthesize how LLMs are integrated into empirical research workflows in software engineering.

In contrast to prior surveys that primarily emphasize task performance and benchmarking~\cite{LLM4SE_LITREVIEW,zhou2022survey}, this review adopts a methodological perspective, examining how LLMs support empirical software engineering activities across the research lifecycle, including study design, data analysis, synthesis, and reporting. Specifically, we (i) classify the empirical research phases in which LLMs are applied, (ii) characterize their functional roles and levels of human involvement, (iii) synthesize the benefits and limitations reported in LLM-assisted empirical work, and (iv) assess reproducibility implications by analyzing how studies document the configurations and conditions under which LLMs are used.

By consolidating fragmented evidence across venues and empirical contexts, this review complements existing performance-oriented surveys and contributes a methodological perspective aimed at supporting rigorous, transparent, and responsible use of LLMs in empirical software engineering research.

\section{Conclusion}
\label{sec:conclusion}

This paper presented a systematic literature review of how LLMs are used to support ESE research tasks. To the best of our knowledge, this is the first systematic literature review to specifically investigate the use of LLMs in supporting empirical software engineering research tasks, providing a consolidated view of their roles, applications, and methodological implications.

By analyzing 50 studies, we provided a comprehensive characterization of the empirical research phases and tasks supported by LLMs, the roles they assume and their levels of human involvement, the benefits and limitations reported in the literature, and the extent to which current studies enable reproducibility through transparent reporting.

Our findings show that LLMs are predominantly applied in structured, data-intensive tasks, particularly within mining software repositories and controlled experiments. Their use is strongly associated with roles such as classification, filtering, and evaluation, and is largely operationalized through automation. While this enables efficiency gains and scalable analysis, it also reflects a constrained form of adoption, in which LLMs are primarily used in contexts where their behavior can be more easily bounded and evaluated. In contrast, their use in tasks requiring deeper reasoning, contextual interpretation, or human-centered decision support remains limited.

From a methodological perspective, our results highlight significant gaps in transparency and reproducibility. Key details required to understand and replicate LLM-assisted analyses, such as prompts, configuration parameters, and validation procedures, are inconsistently reported. This introduces a hidden layer of methodological complexity that is often not adequately documented, limiting the interpretability and replicability of empirical findings.

These findings indicate that LLM-assisted ESE research is still in a formative stage. While LLMs are increasingly adopted as research instruments, their integration remains largely task-specific, automation-centered, and insufficiently standardized. Advancing this area will require more rigorous evaluation of LLM performance within specific empirical contexts, broader exploration of human-AI collaborative workflows, and stronger adherence to reporting practices that ensure transparency and reproducibility.

\section*{Data Availability}

The data underlying this study are publicly available in a GitHub repository \cite{replicationPackage}. This includes the full dataset used in the analysis, comprising the extracted information from the selected studies and the structured coding results presented in this paper. 

\bibliographystyle{ACM-Reference-Format}
\bibliography{references}

\end{document}